\documentclass[preprint,12pt]{elsarticle}
\usepackage{booktabs}
\usepackage{nomencl}
\makenomenclature

\usepackage{amssymb}
\usepackage{amsmath}
\usepackage{cleveref}

\usepackage{outlines}

\journal{Acta materialia}
\begin{document}

\begin{frontmatter}



\title{A GND-Based back stress Model for Reverse Loading in Metal Sheets with consideration of GNB} %% Article title


\address[snu-mse] {Department of Materials Science and Engineering \& RIAM, Seoul National University, 1 Gwanak-ro, Gwanak-gu, Seoul 08826, Korea}
\address[IMS] {Materials Processing Research Division, Korea Institute of Materials Science, Changwon 51508, Republic of Korea}
\address[hyundai]{Materials Research \& Engineering Center, Hyundai Motor Company, 37 Cheoldobangmulgwan-ro, Uiwang-si, Gyeonggi-do 16082, Republic of Korea }

\author[snu-mse]{Gyu-Jang Sim}
\author[snu-mse]{Jehyun You}
\author[IMS]{SeongHwan Choi}
\author[hyundai]{Youngjae Kim} 
\author[hyundai]{Chung An Lee}
\author[hyundai]{Hyunki Kim}
\author[hyundai]{Donghwan Noh}
\author[snu-mse]{Myoung-Gyu Lee\corref{cor1}}

\cortext[cor1]{corresponding author}
\ead{myounglee@snu.ac.kr}

% \affiliation{organization={},%Department and Organization

%% Abstract
\begin{abstract}
Accurate prediction of springback and formability in sheet metal forming requires understanding reverse loading behavior under complex loading path changes, such as tension followed by compression. However, for ultra-thin sheets experimental characterization of such behavior is difficult due to compressive instability like plastic buckling. This study presents a crystal plasticity finite element method (CPFEM) incorporating a physically motivated back stress model based on geometrically necessary dislocations (GNDs) and boundaries (GNBs). The model captures grain size effects, including the Hall-Petch and Bauschinger effects, through a single grain size-dependent back stress parameter, enabling reverse loading prediction using only tensile data from specimens with different grain sizes. The back stress parameter was calibrated by fitting tensile stress-strain curves from two microstructures—one as-received and one annealed. Without using Tension-Compression (T-C) data for calibration, the model accurately predicted reverse loading behavior in low-carbon steel (0.64 mm thick) and Tension-Bending (T-B) responses in ultra-thin SUS316 (0.083 mm thick) when the developed theory was incorporated to an upscaled anisotropic hardening model. Identifiability analysis confirmed that the model parameters are uniquely determined by the available data. This physically interpretable framework provides an efficient and robust means to predict reverse loading in thin metal sheets, overcoming experimental limitations.

\end{abstract}


\begin{highlights}
\item CPFEM-based method predicts metal sheet behavior under load reversal.
\item Bauschinger effect was predicted quantitatively using two tensile tests from samples with different grain sizes.
\item Key parameters identified from tensile tests on samples with varied grain sizes.
\item Validated by tension-compression and tension-bending experiments.
\end{highlights}

%% Keywords

\end{frontmatter}



\section*{Nomenclature}

\begin{tabular}{ll}
$\varepsilon_{ijk}$ & Levi-Civita permutation symbol \\
$\otimes$           & Dyadic product \\
$\mathbf{I}$        & Second-order identity tensor \\

$G$                 & Shear modulus \\
$\mathbb{C}^e$      & Fourth-order elasticity tensor \\
$\boldsymbol{\sigma}$ & Cauchy stress tensor \\
$\mathbf{S}$        & Piola--Kirchhoff stress tensor \\
$\mathbf{E}^*$      & Green--Lagrange strain tensor \\

$\mathbf{F}$        & Deformation gradient \\
$\mathbf{F}^e$      & Elastic deformation gradient \\
$\mathbf{F}^p$      & Plastic deformation gradient \\
$\mathbf{L}^p$      & Velocity gradient of plastic deformation \\

$k$                 & Boltzmann constant \\
$T$                 & Absolute temperature \\

$\dot{\gamma}^a$    & Slip rate on slip system $a$ \\
$\mathbf{s}_0^a$    & Unit vector in slip direction (reference configuration) \\
$\mathbf{n}_0^a$    & Unit normal to slip plane of slip system $a$ (reference configuration) \\
$\tau^a$            & Resolved shear stress on slip system $a$ \\
$\tau_{\text{back}}^a$ & Back stress on slip system $a$ \\
$\dot{\gamma}_0$    & Reference shear rate \\
$m$                 & Strain-rate sensitivity exponent \\
$\rho_M^a$          & Mobile dislocation density on slip system $a$ \\
$b$                 & Burgers vector magnitude \\
$v^a$               & Average velocity of mobile dislocations on slip system $a$ \\
$\nu_0$             & Attempt frequency of dislocation glide \\
$s^a$               & Slip resistance on slip system $a$ \\
$\rho^a$            & Dislocation density on slip system $a$ \\
$k_1, k_2$          & Kocks--Mecking law parameters \\
$\gamma^a$          & Slip amount on slip system $a$ \\

$\boldsymbol{\sigma}^{\text{dis}}$ & Stress from dislocations \\
$\boldsymbol{\sigma}^{\text{ext}}$ & Applied external stress \\
$\boldsymbol{\sigma}^{\text{img}}$ & Image stress \\
$\boldsymbol{\sigma}^{\text{FE}}$  & Finite element model stress \\
$\boldsymbol{\varepsilon}^{p}$     & Plastic strain \\
$\overset{\nabla}{\boldsymbol{\sigma}}$ & Jaumann rate of Cauchy stress \\

% next page
\end{tabular}
\begin{tabular}{ll}

$r_{PK}$            & Scaling parameter for Peach--Koehler force \\

$\rho_{t,e}^a$      & Total edge dislocation density on slip system $a$ \\
$\rho_{t,s}^a$      & Total screw dislocation density on slip system $a$ \\
$\rho_{\text{GND,e}}^a$ & GND edge dislocation density on slip system $a$ \\
$\rho_{\text{GND,s}}^a$ & GND screw dislocation density on slip system $a$ \\
$\rho_{\text{SSD,e}}^a$ & SSD edge dislocation density on slip system $a$ \\
$\rho_{\text{SSD,s}}^a$ & SSD screw dislocation density on slip system $a$ \\
$(\rho_{\text{SSD}}^a)_0$ & Initial SSD density on slip system $a$ \\
$\alpha$            & Geometric factor \\
$\dot{\rho}_{KM}^a$ & Dislocation generation rate by KM law on slip system $a$ \\
$\boldsymbol{\Lambda}$ & Incompatibility tensor \\
$\mathbf{l}^a$      & Dislocation line vector on slip system $a$ \\

$B^a$               & Magnitude of superdislocation Burgers vector on slip system $a$ \\
$L^a$               & Element length in slip direction \\
$V^a$               & Element volume \\

$p$                 & Number of nodes \\
$q$                 & Number of integration points \\
$[\mathbf{T}]_{3\times p}$ & Gradient operator \\
$[\mathbf{J}]_{3\times 3}$ & Jacobian matrix \\
$[\mathbf{N}]_{1\times p}$ & Shape function in isoparametric space \\
$[\mathbf{N}]_{p\times q}$ & Shape functions at integration points \\
$[\nabla \mathbf{N}]_{3\times p}$ & Shape function derivatives in isoparametric space \\
$[\mathbf{x}]_{p \times 3}$ & Physical nodal coordinates \\
$\mathbf{G}$        & Gradient tensor \\

$h$                 & Average distance between activated slip systems \\
$D$                 & Average grain size \\
$\gamma^{\text{sat}}$ & Slip saturation threshold \\
$\zeta$             & Backstress parameter \\

$q$                 & Yield surface distortion exponent \\
$k$                 & Microstructure deviator evolution rate \\
$k_g$               & Distortion evolution rate \\
$\tau_g$            & Distortion saturation value \\
$\gamma_g$          & Distortion saturation exponent \\
$k_p$               & Permanent softening rate \\
$\tau_p$            & Permanent softening saturation value \\
\end{tabular}

\section{Introduction}
\label{introduction}

Understanding the mechanical behavior of metal sheets under load reversal is crucial for accurately predicting springback during the sheet metal forming process. 
In forming processes, sheet metals undergo significant loading path changes, including tension followed by compression, or vice versa. 
The stress-strain curve under load reversal conditions can significantly affect the final shape of the formed part~\cite{WAGONER20133}.
A key aspect to consider is the anisotropic hardening behavior, such as the Bauschinger effect~\cite{bauschinger1881} and other transient behaviors, which can influence the magnitude of springback by a factor of two~\cite{GengWagoner2000}. 
The Bauschinger effect refers to the phenomenon where the yield stress in the reverse loading direction is lower than the yield stress in the forward loading direction.

The most common way to acquire the mechanical behavior is through in-plane Tension-Compression (T-C) testing, where the gauge section of a specimen is subjected to uniaxial tensile strain followed by uniaxial compressive strain. 
The challenge is that the thinner the sheet, the more prone it is to buckling under compression. 
For example, recent fuel cells require thinner metal bipolar plates to reduce weight and improve efficiency~\cite{LI2005359,BONG20121}, and battery pouch designs demand ultra-thin sheets to optimize space and weight~\cite{MOON2023108601,MOON2025237746}, leading to challenges in accurately predicting springback and formability due to the complex loading paths experienced during forming processes.
To obtain the T-C behavior without buckling, anti-buckling devices have been used during compression~\cite{BOGER20052319}.

However, for extremely thin sheets, typically with a thickness lower than 0.1 mm, even anti-buckling devices become ineffective.
To address these challenges, various indirect experimental approaches with macroscopic constitutive models have been developed. Experimental processes include forward-reverse simple shear tests~\cite{HU1992839,CHOI2015144}, bending-reverse bending tests~\cite{YOSHIDA1998237,GENG2002743}, and Tension-Bending (T-B) tests~\cite{ZANG201484,CHOI2019428}. In these tests, load reversal with complex stress states, other than uniaxial tension-compression, is applied to the specimen. The resulting force-displacement curve data or final specimen shapes are used to calibrate the parameters of constitutive models that describe the macroscopic stress-strain response under load reversal. Commonly used models for capturing the Bauschinger effect and related phenomena include the Yoshida-Uemori (Y-U) model~\cite{YOSHIDA2002661,YOSHIDA2002633}, the Homogeneous Anisotropic Hardening (HAH) model~\cite{BARLAT20111309}, and their modified versions~\cite{BARLAT2013130,CHOI2025109856}. The parameters of the constitutive models are typically identified using optimization techniques, such as least-squares fitting to match the experimental results.

On the other hand, beyond the phenomenological models, mesoscale frameworks such as Visco-Plastic Self-Consistent (VPSC)~\cite{LEBENSOHN19932611} and Crystal Plasticity Finite Element Method (CPFEM)~\cite{Kalidindi1992} have been developed to capture mesoscopic plasticity at the grain scale. Recent studies have introduced microstructural origins of the Bauschinger effect, providing insights into microstructure-property relationships. For example, Beyerlein et al.~\cite{BEYERLEIN2007640} introduced a back stress formulation within the VPSC framework, and Wen et al.~\cite{WEN2016305} further extended VPSC to incorporate back stress formulations for cyclic loading. In the CPFEM framework, Li et al.~\cite{LI2014174} developed a texture-based Representative Volume Element (RVE) to predict the Bauschinger effect in rolled polycrystalline aluminum alloy 7075 under cyclic loading. Similarly, Cruzado et al.~\cite{CRUZADO2017148} employed back stress in RVE-based CPFEM to capture the stress-strain behavior of nickel-based superalloys under cyclic loading.
The models successfully reproduced the Bauschinger effect within the VPSC and CPFEM frameworks. 
However, these models require experimental data with load reversal for calibration and rely on somewhat phenomenological terms, such as the Armstrong-Frederick (AF)~\cite{Frederick2007} or Ohno-Wang (OW) kinematic hardening models~\cite{OHNO1993391}.

Meanwhile, other CPFEM models have been developed to both reproduce and explain the microstructural origins of the Bauschinger effect.
Lim et al.~\cite{LIM20111328} proposed a Super-Dislocation (SD) concept that incorporates elastic dislocation interactions by superimposing Peach-Koehler (PK) forces between integration points. 
This approach captures the inter-granular elastic interactions between dislocations and the back stress effects arising from dislocation structures. The model was further extended by Zhou et al.~\cite{ZHOU2020104178} to include geometrically necessary dislocation (GND) density and their elastic interactions, referred to as the GM model. 
The approach successfully captured stress-strain hysteresis under tension loading-unloading-reloading and was further used to explain yield point phenomena (YPP)~\cite{10.1115/1.4051855}. 
However, to the best of our knowledge, such PK force-based back stress models have not been compared with experimental load reversal behavior, such as T-C tests, and their computational cost is high, making them impractical for predicting macroscopic load reversal behavior.
Similarly, higher-order strain gradient models—which compute back stress based on the gradient of GND density~\cite{KURODA20082573,Ma2014,ZHANG2023103553}—have also not been applied to predict or validate macroscopic load reversal behavior, such as T-C tests. Instead, their application has mostly been limited to thin films or microstructures containing only a few grains.

On the other hand, simpler back stress models accounting for GND density have been proposed to capture the Bauschinger effect.
For example, Kapoor et al.~\cite{KAPOOR2018447} defined the back stress as
\begin{equation}
\label{eq:back stress_KAPOOR2018447}
\tau_{back}^a = K G b \sqrt{\rho_{GND}^a}\;,
\end{equation}

where \(\tau_{back}^a\) is the back stress on slip system \(a\), \(K\) is a material constant, \(G\) is the shear modulus, \(b\) is the Burgers vector, and \(\rho_{GND}^a\) is the GND density on slip system \(a\). 
This GND-induced back stress model was further examined by Bandyopadhyay~\cite{BANDYOPADHYAY2021102887}, who validated it against high-energy X-ray diffraction microscopy (HEDM) experiments that capture grain-resolved stress responses under cyclic loading. 
However, these existing models lack a clear explanation of how the back stress expressions are derived from microstructural mechanisms. 

In summary, macroscopic models have reasonably well reproduced the Bauschinger effect~\cite{YOSHIDA2002661,YOSHIDA2002633,BARLAT20111309,BARLAT2013130,CHOI2025109856}, but they still lack explanations based on physically realistic microstructural characteristics.
Particularly, CPFEM or VPSC models that utilize phenomenological back stress terms have captured the Bauschinger effect with some degree of microstructural explanation~\cite{BEYERLEIN2007640,WEN2016305,LI2014174,CRUZADO2017148}, but these terms are not derived from an understanding of physical mechanisms. 
CPFEM models that take into account elastic dislocation interactions, such as the SD~\cite{LIM20111328}, general mesoscale (GM) model~\cite{ZHOU2020104178}, or higher-order strain gradient models~\cite{KURODA20082573,Ma2014,ZHANG2023103553}, aim to explain the microstructural origins of the Bauschinger effect. However, the high computational cost and the lack of validation against load reversal experiments limit their application to macroscopic load reversal behavior.
GND density-based back stress models~\cite{KAPOOR2018447} have been proposed to capture the Bauschinger effect and have been well validated by experiments involving load reversal, but they also lack a clear mechanistic understanding linking dislocation structures to back stress expressions.

In this study, we address this gap in the literature by proposing a methodology that utilizes a physically motivated back stress model to predict the mechanical behavior of metal sheets under load reversal conditions. A key motivation behind this work is the belief that if the model is based on physical mechanisms and its parameters are uniquely identifiable, it should be capable of predicting the material's mechanical response under load reversal conditions, even without using data from T-C or T-B tests for calibration.

\begin{figure}[!htbp]
\centering
\includegraphics[width=\textwidth]{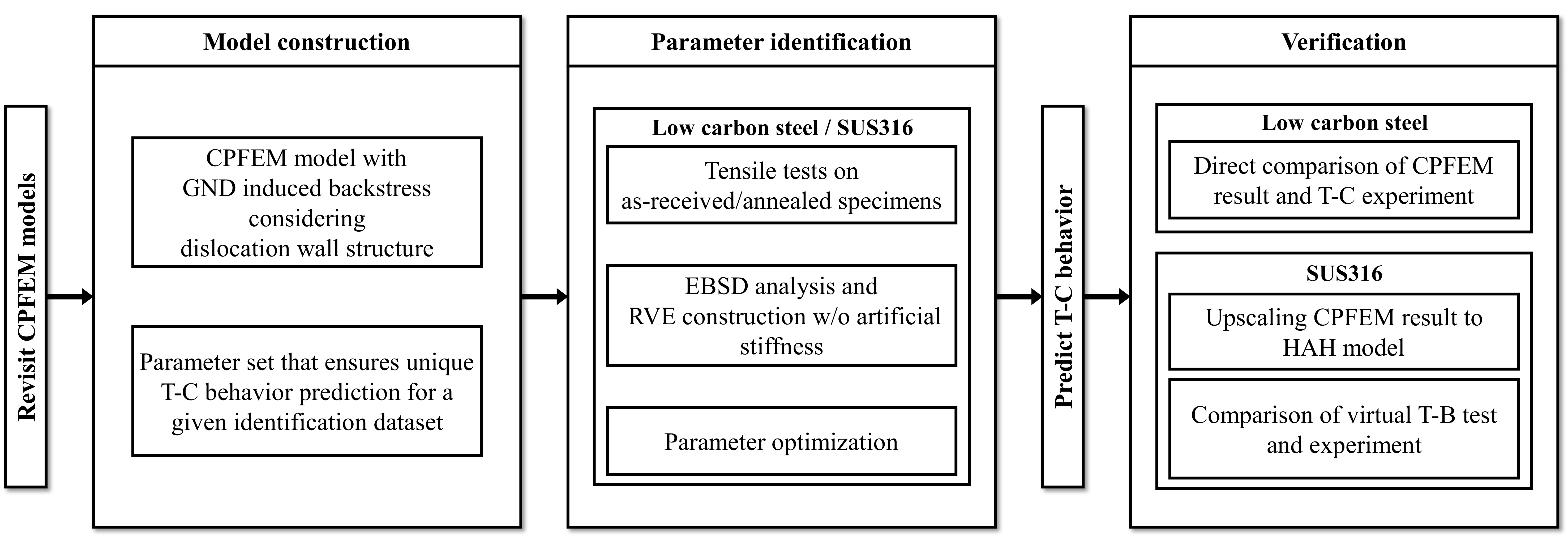}
\caption{Overall flowchart of this study.}
\label{fig:Overview}
\end{figure}

In other words, physics-based models should be able to extrapolate stress-strain curves under loading conditions not directly included in the calibration data, unlike phenomenological or data-driven models. 
To examine this idea, we proceed as shown in \cref{fig:Overview}. 
Our methodology is applied to two commonly used metal sheets. 
The first is a low-carbon steel sheet with a ferritic microstructure and a thickness of 0.64~mm. 
In this case, an anti-buckling device can be applied for direct T-C testing, enabling experimental validation. 
The second is SUS316 sheet with a thickness of 0.083~mm, where an anti-buckling device cannot be readily effective in T-C tests due to its extremely thin gauge, which is prone to buckling during compressive loading.
To avoid instability in the T-C test for ultra-thin sheet metal, an alternative T-B test~\cite{ZANG201331} was conducted, where the outer and inner surfaces of the bent sample experience tensile-tensile and tensile-compressive strains, respectively—thus enabling indirect validation of T-C behavior. 
To compare CPFEM predictions with macroscopic T-B experiments, the CPFEM RVE results were upscaled using the HAH model. 
The HAH parameters were calibrated to match the CPFEM T-C response and then used in finite element simulations of the T-B test geometry for direct comparison with experimental force-displacement data.

This paper is organized as follows. 
Section \ref{Revisiting CPFEM models} reviews the kinematics, dislocation density-based constitutive laws in CPFEM, and the SD model concept.
Issues in the previously reported SD model are addressed, which are enhanced by the proposed models in this study.
This section also describes how the RVE and Periodic Boundary Conditions (PBCs) are established.
Section \ref{GND-based back stress model} introduces the GND-based back stress model, derived from the concept of dislocation wall structures and incorporating both GND density and slip system orientation. 
Section \ref{Experimental} outlines the experimental procedures, including the T-C and T-B tests for low-carbon steel and SUS316, respectively. 
Section \ref{Results} presents the results of the CPFE simulations, including parameter calibration and validation using experimental data. 
Section \ref{Discussion} examines the uniqueness of the predicted stress-strain response under load reversal conditions, demonstrating that the proposed methodology yields a single solution. 
Additional virtual experiments are also discussed.
Finally, conclusions are summarized in Section \ref{Conclusions}.
All simulations were performed in ABAQUS~\cite{abaqus2024}, using the UMAT and UEXTERNALDB user subroutines.

% ACRONYM CHECK

% \section{Review of SD model}
% \label{Review of SD model}

\section{Revisiting CPFEM and RVE construction}
\label{Revisiting CPFEM models}

% In this section, we review the basic kinematics of CPFEM, dislocation density based CPFEM models, and the SD model. The SD model is revisited to address the issues encountered in our own implementation. We also discuss the RVE construction and periodic boundary condition (PBC) used in this study.

% \subsection{CPFEM}
% \label{CPFEM}

The kinematics of CPFEM follow conventional formulations, where the total deformation gradient $\mathbf{F}$ is multiplicatively decomposed into elastic and plastic components~\cite{ASARO1983}:

\begin{equation}
\label{eq:deformation_gradient}
\mathbf{F} = \mathbf{F}^e \mathbf{F}^p.
\end{equation}

Here, $\mathbf{F}^e$ denotes the elastic deformation gradient, including both elastic distortion and lattice rotation, while $\mathbf{F}^p$ represents the plastic deformation gradient, which captures the plastic slip from dislocation glide. 
To incorporate the slip behavior of individual slip systems into the continuum framework, the velocity gradient of the plastic deformation, $\mathbf{L}^p$, is expressed as the sum of the slip contributions from all active slip systems as:
\begin{equation}
\label{eq:plastic_velocity_gradient}
\mathbf{L}^p = \dot{\mathbf{F}}^p{\mathbf{F}^p}^{-1} = \sum_a \dot{\gamma}^a \mathbf{s}_0^a \otimes \mathbf{n}_0^a,
\end{equation}

where $\dot{\mathbf{F}}^p$ is the time derivative of the plastic deformation gradient, $\dot{\gamma}^a$ is the slip rate on slip system \(a\), $\mathbf{s}_0^a$ is the unit vector in the slip direction, and $\mathbf{n}_0^a$ is the unit normal to the slip plane of slip system \(a\). 
Both $\mathbf{s}_0^a$ and $\mathbf{n}_0^a$ are defined in the reference configuration.
The dyadic product $\mathbf{s}_0^a \otimes \mathbf{n}_0^a$ defines the Schmid tensor associated with each slip system.
The slip rate can be expressed as a function of the resolved shear stress acting on slip system \(a\). 
In this study, we utilize the power-law formulation:

\begin{equation}
\label{eq:slip_rate}
\dot{\gamma}^a = \dot{\gamma}_0 \left< \frac{\tau^a - \tau_{back}^a}{s^a} \right>^{1/m} \operatorname{sgn}(\tau^a - \tau_{back}^a),
\end{equation}

where $\tau^a$ is the resolved shear stress on slip system $a$, and $\tau_{back}^a$ denotes the back stress on the same system. The term $s^a$ corresponds to the slip resistance, also referred to as the critical resolved shear stress. 
$\dot{\gamma}_0$ is the reference shear rate, and $m$ is the strain rate sensitivity exponent. 
The operator $\operatorname{sgn}$ is 1 if the argument is positive, -1 if negative, and 0 if zero. 
The operator $\left< \cdot \right>$ denotes the Macaulay bracket, which is defined as $\left< x \right> = \max(0,x)$, ensuring that slip does not occur when the resolved shear stress is less than the back stress.

Once the plastic part of the deformation gradient is determined based on the slip activity of each slip system by \cref{eq:plastic_velocity_gradient,eq:slip_rate}, the elastic deformation gradient $\mathbf{F}^e$ is determined by \cref{eq:deformation_gradient}, assuming $\mathbf{F}$ is known, as is typical during the stress update step in standard FE algorithms. 
Then, the stress can be evaluated using generalized Hooke's law applied to the elastic part of the deformation gradient. The second Piola-Kirchhoff stress tensor $\mathbf{S}$ is obtained as:

\begin{equation}
\label{eq:stress_tensor}
\mathbf{S} = \mathbb{C}^e : \mathbf{E}^* = \det(\mathbf{F}^e) \mathbf{F}^{e^{-1}} \boldsymbol{\sigma} \mathbf{F}^{e^{-T}},
\end{equation}

where $\mathbb{C}^e$ is the fourth-order elasticity tensor, and $\boldsymbol{\sigma}$ is the Cauchy stress tensor. For cubic crystal structures, in Voigt notation, the elasticity tensor $\mathbb{C}^e$ contains only three independent components: $C_{11}$, $C_{12}$, and $C_{44}$. 
The components satisfy $C_{11} = C_{22} = C_{33}$, $C_{12} = C_{21} = C_{13} = C_{31} = C_{23} = C_{32}$, and $C_{44} = C_{55} = C_{66}$, while all other components are zero~\cite{nye1985}. The Green--Lagrange strain tensor $\mathbf{E}^*$ is computed from the elastic deformation gradient $\mathbf{F}^e$ as:

\begin{equation}
\label{eq:green_lagrange_strain}
\mathbf{E}^* = \frac{1}{2} \left( \mathbf{F}^{e^T} \mathbf{F}^e - \mathbf{I} \right),
\end{equation}

where $\mathbf{I}$ is the identity tensor. 

The formulations presented in \cref{eq:deformation_gradient,eq:plastic_velocity_gradient,eq:slip_rate,eq:stress_tensor,eq:green_lagrange_strain} represent the fundamental kinematics and constitutive laws of CPFEM, which were first implemented in the stress update algorithm within the FEM framework by Kalidindi et al.~\cite{Kalidindi1992}.

On the other hand, the power-law slip rate formulation (\cref{eq:slip_rate}) is sometimes replaced by the physically motivated Orowan equation~\cite{Orowan1934} in other CPFEM implementations~\cite{MA20043603,MIN2024104049}:
\begin{align}
     \dot{\gamma}^a = \rho_M^a b v^a = \rho_M^a b \nu_0 \exp\left(-\frac{Q(\tau^a)}{kT}\right),
\end{align}
where $\rho_M^a$ is the mobile dislocation density, $b$ is the Burgers vector magnitude, and $v^a$ is the average velocity of mobile dislocations on slip system $a$, expressed as $v^a = \nu_0 \exp\left(-\frac{Q(\tau^a)}{kT}\right)$, with $\nu_0$ being the attempt frequency of dislocation glide, $Q(\tau^a)$ the stress-dependent activation energy, $k$ the Boltzmann constant, and $T$ the absolute temperature. 
While the Orowan equation is physically motivated, Li et al.~\cite{LI2021102921} demonstrated its compatibility with the power-law formulation, which is used here for simplicity.

The details of the CPFEM formulation are further developed in the expressions for slip resistance $s^a$ and back stress $\tau_{back}^a$. In the following subsections, we review how dislocation densities are incorporated into the dislocation density-based CPFEM model (\cref{Dislocation density based CPFEM model}) to describe $s^a$, and revisit the SD model (\cref{SD model}), which introduces elastic dislocation interactions for $\tau_{back}^a$. Finally, we discuss RVE construction and periodic boundary conditions, highlighting important considerations (\cref{RVE and PBC}).

\subsection{Dislocation density based CPFEM}
\label{Dislocation density based CPFEM model}

To incorporate dislocation density as a state variable into the CPFEM framework, researchers~\cite{LEE2010925, AOYAGI201313, REZVANIAN200880, SHANTHRAJ20117695, ZIAEI2016435} have expressed the slip resistance $s^a$ as a function of dislocation density, following the classical form of Taylor hardening~\cite{TAYLOR1934}, as introduced by Franciosi et al.~\cite{FRANCIOSI1980273}. Here, the model constructed by Lee et al.~\cite{LEE2010925} is reviewed with some modifications in notation for consistency with the rest of the paper. 

Considering the angle between slip systems, the forest dislocation density is defined as:

\begin{equation}
\label{eq:forest_dislocation_density}
\rho_{f}^a=\sum_{b} \left|\boldsymbol{\mathrm{n}}^{a} \cdot \boldsymbol{\mathrm{t}}^{b} \right| \rho^b,
\end{equation}

where $\boldsymbol{\mathrm{n}}^a$ is the unit normal vector to the slip plane of slip system \(a\), $\boldsymbol{\mathrm{t}}^b$ is the line direction of dislocations on slip system \(b\), and $\rho^b$ is the dislocation density on slip system \(b\). 

Based on the concept of forest dislocation density, the average spacing between dislocations that hinder slip on slip system \(a\) is assumed to be inversely proportional to the square root of the forest dislocation density \(\rho_{f}^a\). Since slip resistance is inversely related to this spacing at the mesoscale, the slip resistance \(s^a\) can be expressed as:
\begin{equation}
\label{eq:s}
s^{a} = G b \sqrt{\rho_f^a},
\end{equation}

where \(G\) is the shear modulus, and \(b\) is the magnitude of the Burgers vector, following the Taylor hardening law.

Then, the evolution of dislocation density is statistically described by the Kocks-Mecking (KM) evolution law~\cite{KOCKS2003171}:

\begin{equation}
\label{eq:KM_evolution_law}
\dot \rho^{a} = \left(k_{1} \frac{\sqrt{\rho^{a} } }{b }-k_{2} \rho ^{a}\right)|\dot \gamma ^{a} |,
\end{equation}

where $k_1$ and $k_2$ are material constants, $b$ is the magnitude of the Burgers vector, $\dot \rho^{a}$ is the time derivative of dislocation density on slip system \(a\), and $|\dot \gamma ^{a}|$ is the absolute value of the slip rate on slip system \(a\). The first term in \cref{eq:KM_evolution_law}, $k_1 {\sqrt{\rho^{a}}}/{b}$, represents the generation of dislocations, while the second term, $-k_2 \rho^{a}$, represents the annihilation of dislocations.
The model described by \cref{eq:forest_dislocation_density,eq:s,eq:KM_evolution_law} is named Single Crystal Constitutive equations based on Dislocation density (\textbf{SCCE-D}) model, and the model well captured flow curves of FCC and BCC single crystals, polycrystal stress-strain curves under tensile loading, and texture evolution in polycrystalline materials~\cite{LEE2010925}. $\tau_{back}^a$ is not included in this model.

\subsection{Super-Dislocation (SD) model from Discrete Dislocation Dynamics (DDD) perspective}
\label{SD model}

To incorporate long-range elastic dislocation interactions into CPFEM, Lim et al.~\cite{LIM20111328} introduced the SD model. In this model, each integration point is treated as a superdislocation carrying the total Burgers vector accumulated from the local dislocation density, which is defined as:
\begin{equation}
\label{eq:superdislocation_burgers_vector}
B^{a} = n^{a} b = \frac{\rho^{a} V^{a} b}{L^{a}},
\end{equation}
where $B^a$ is the magnitude of the superdislocation's Burgers vector on slip system $a$, $\rho^a$ is the dislocation density, $V^a$ is the volume of the element, and $L^a$ is the element length through the slip direction. The PK force between two superdislocations located at integration points $i$ and $j$ is computed using analytical expressions for the stress field induced by a dislocation in an infinite elastic medium~\cite{anderson2017theory}. The back stress $\tau_{\text{back}}^a$ at point $i$ is then evaluated by summing these forces:
\begin{equation}
\label{eq:back stress_from_PK_force}
\left(\tau_{\text{back}}^a\right)_i = \frac{1}{b} \sum_{j \neq i} F_{ij}^a,
\end{equation}

where \(F_{ij}^a\) is the PK force acting on superdislocation \(i\) due to superdislocation \(j\) on slip system \(a\). 
While this approach has intuitive appeal, it introduces redundancy when viewed from the perspective of recent developments in Discrete Dislocation Dynamics (DDD) and their coupling with continuum frameworks.
DDD models simulate the motion of individual dislocations and their interactions as discrete entities, enabling detailed resolution of plasticity at submicron scales~\cite{BulatovCai2006}. However, DDD does not directly solve boundary value problems (BVPs). Instead, it assumes a uniform external stress field and uses analytical expressions for dislocation stress under the assumption of an infinite elastic medium.
\begin{equation}
\boldsymbol{\sigma} = \boldsymbol{\sigma}^{\text{dis}} + \boldsymbol{\sigma}^{\text{ext}} + \boldsymbol{\sigma}^{\text{img}},
\end{equation}
where $\boldsymbol{\sigma}^{\text{dis}}$ is the stress from dislocations, $\boldsymbol{\sigma}^{\text{ext}}$ is the applied external stress, and $\boldsymbol{\sigma}^{\text{img}}$ is the image stress ensuring traction-free boundaries.

To apply DDD under complex boundary conditions, hybrid models combining DDD with FEM have been proposed. 
The Superposition Method (SPM)~\cite{van_der_Giessen_Needleman_1995} is one such approach. 
In SPM, the dislocation stress $\boldsymbol{\sigma}^{\text{dis}}$ is computed in infinite media and added to the finite-element correction field $\boldsymbol{\sigma}^{\text{FE}}$, which is obtained by solving the boundary value problem.
\begin{equation}
\boldsymbol{\sigma}^{\text{total}} = \boldsymbol{\sigma}^{\text{dis}} + \boldsymbol{\sigma}^{\text{FE}}.
\end{equation}
The dislocation configuration affects the FE results with boundary traction corrections, and the resolved stress field is fed back into DDD for calculating dislocation motion. 
Here, FEM only solves for elastic strain and does not account for plastic strain. 

\begin{figure}[!htbp]
\centering
\includegraphics[width=\textwidth]{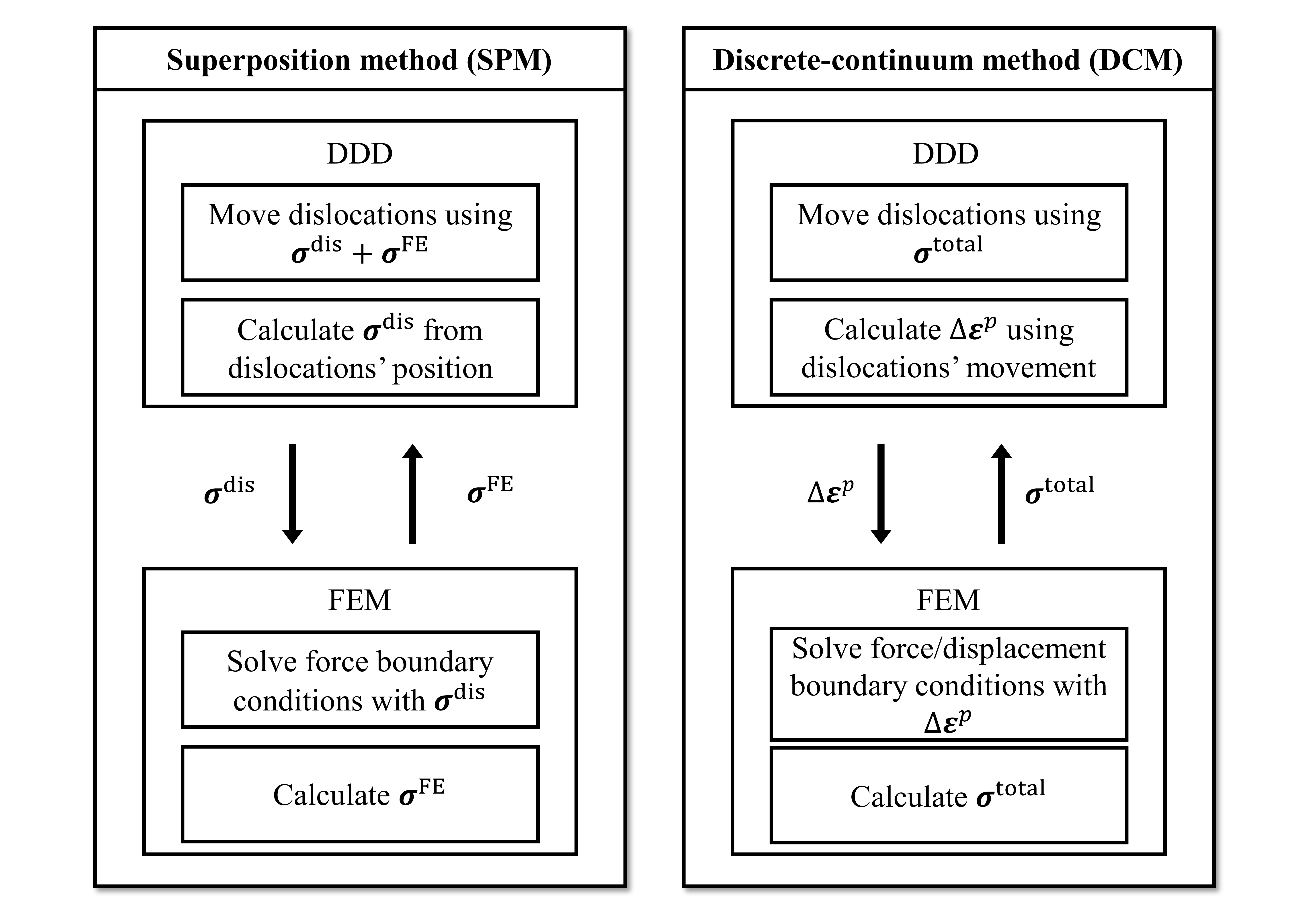}
\caption{Comparison of the SPM and DCM for coupling DDD and FEM. In SPM, dislocation stress is analytically computed and superimposed with FEM correction, while in DCM, plastic strain from dislocation motion is transferred to FEM, which solves for the total stress field.}
\label{fig:DD_FE}
\end{figure}

In contrast, the Discrete-Continuum Model (DCM)~\cite{lemarchand2001, CUI201554, VATTRE2014491} or Defect Dynamics Model (DDM)~\cite{CUONGNGUYEN2025113132} introduces the concept of eigenstrain. The plastic strain $\boldsymbol{\varepsilon}^{p}$, resulting from dislocation glide, is transferred from DDD to FEM as:
\begin{equation}
\overset{\nabla}{\boldsymbol{\sigma}} = \mathbb{C}^e : (\dot{\boldsymbol{\varepsilon}} - \dot{\boldsymbol{\varepsilon}}^p),
\end{equation}
where $\overset{\nabla}{\boldsymbol{\sigma}}$ is the Jaumann rate of Cauchy stress. With given boundary conditions and internal plastic strain, FEM solves the stress field without separately computing dislocation stress and boundary-induced corrections. 
The feedback is two-way: DDD provides $\boldsymbol{\varepsilon}^p$, and FEM provides $\boldsymbol{\sigma}$. 
According to Lemarchand~\cite{lemarchand2001}, supported by subsequent studies~\cite{CUI201554, VATTRE2014491}, this DCM approach inherently reflects the effects of stress fields generated by dislocations without explicitly computing them. It simultaneously solves for the stress field up to the boundaries under the given boundary conditions.
A summary of the differences between SPM and DCM is shown in \cref{fig:DD_FE}.

From the perspective of the FE model, only the analytical dislocation-induced stress is transferred from the DD model to the FE model in SPM. 
In DCM, only the plastic strain generated by dislocation motion is passed from the DD model to the FE model.

Reconsidering the SD model in this context, the issue becomes apparent. Since CPFEM already incorporates the plastic strain through the multiplicative decomposition $\mathbf{F} = \mathbf{F}^e \mathbf{F}^p$ and computes stress accordingly, the additional inclusion of PK forces between superdislocations effectively double-counts dislocation interactions. In other words, it amounts to superimposing SPM over a DCM-consistent framework, which is physically redundant.

\begin{figure}[!htbp]
\centering
\includegraphics[width=\textwidth]{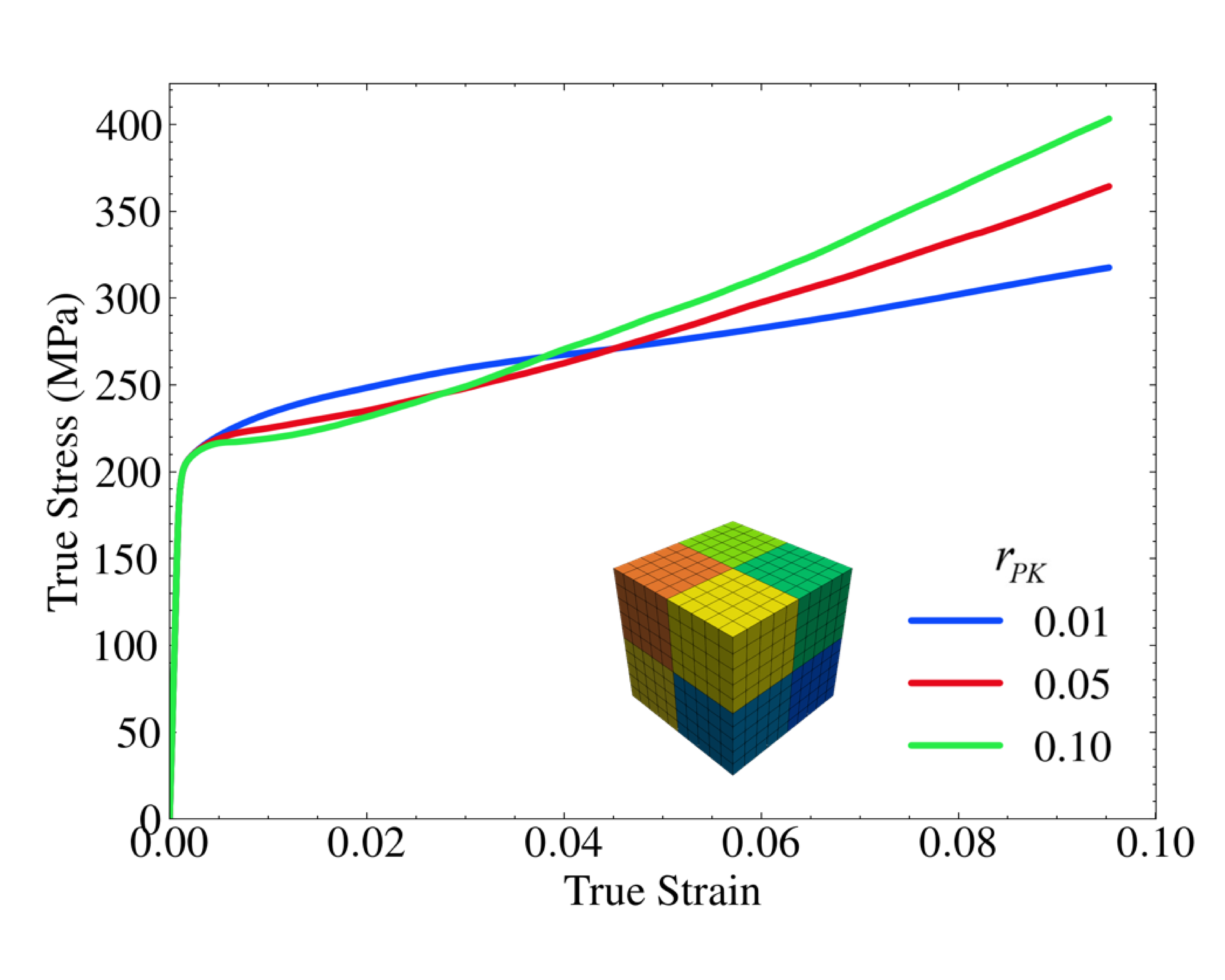}
\caption{Stress-strain curves of the RVE simulation with PK forces. As the scaling parameter ($r_{PK}$) for PK force increases, the stress-strain curve becomes concave downward at higher strains, diverging from experimental results. This behavior is due to the cyclic feedback loop between slip-induced PK forces and further slip, leading to double counting of elastic dislocation interactions.}
\label{fig:HLIM_model_sscurves}
\end{figure}

In our revisiting implementation of the model, simulations of tensile loading on RVEs produced abnormal responses where the stress-strain curve became concave downward at higher strains (\cref{fig:HLIM_model_sscurves}), diverging from common experimental results. In this implementation, the PK force scaling parameter \(r_{PK}\) was applied to \cref{eq:back stress_from_PK_force} as a multiplier to control the magnitude of the PK force. As \(r_{PK}\) increases from 0.01 to 0.10, the PK force becomes more significant, leading to a concave downward curve at higher strains. This occurred because the slip-induced PK force led to further slip, resulting in a cyclic feedback loop and a double-counting issue.

\begin{figure}[!htbp]
\centering
\includegraphics[width=\textwidth]{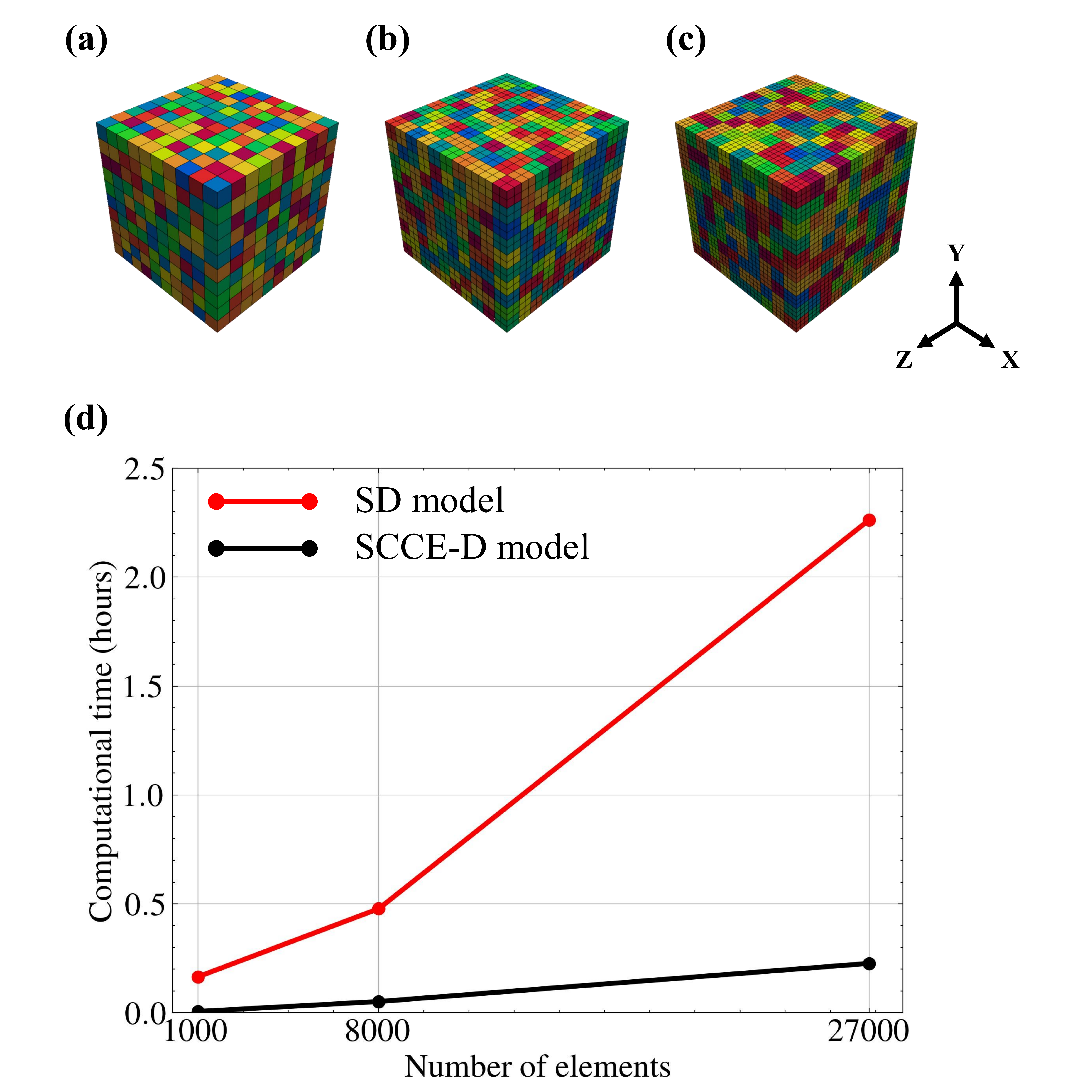}
\caption{Computational time (in seconds) required for RVE simulations up to 10\% engineering strain under 
PBC. As the number of elements increases—$10^3$, $20^3$, and $30^3$ elements for (a), (b), and (c), respectively—the computational time drastically increases compared to the SCCE-D model as shown in (d).}
\label{fig:HLIM_model_computational_time}
\end{figure}

Moreover, PK forces between superdislocations frequently exceeded the critical resolved shear stress (CRSS), requiring artificial stabilizers for numerical convergence. Additionally, computing PK forces between all integration points incurs an \( O(n^2) \) computational cost, where \(n\) is the number of integration points per grain, making the method impractical for parameter calibration based on experimental data. \Cref{fig:HLIM_model_computational_time} shows the computational time for the RVE simulation with PK forces, which increases quadratically with the number of integration points. This makes the method impractical for parameter calibration based on experimental data.

\begin{figure}[!htbp]
\centering
\includegraphics[width=0.8\textwidth]{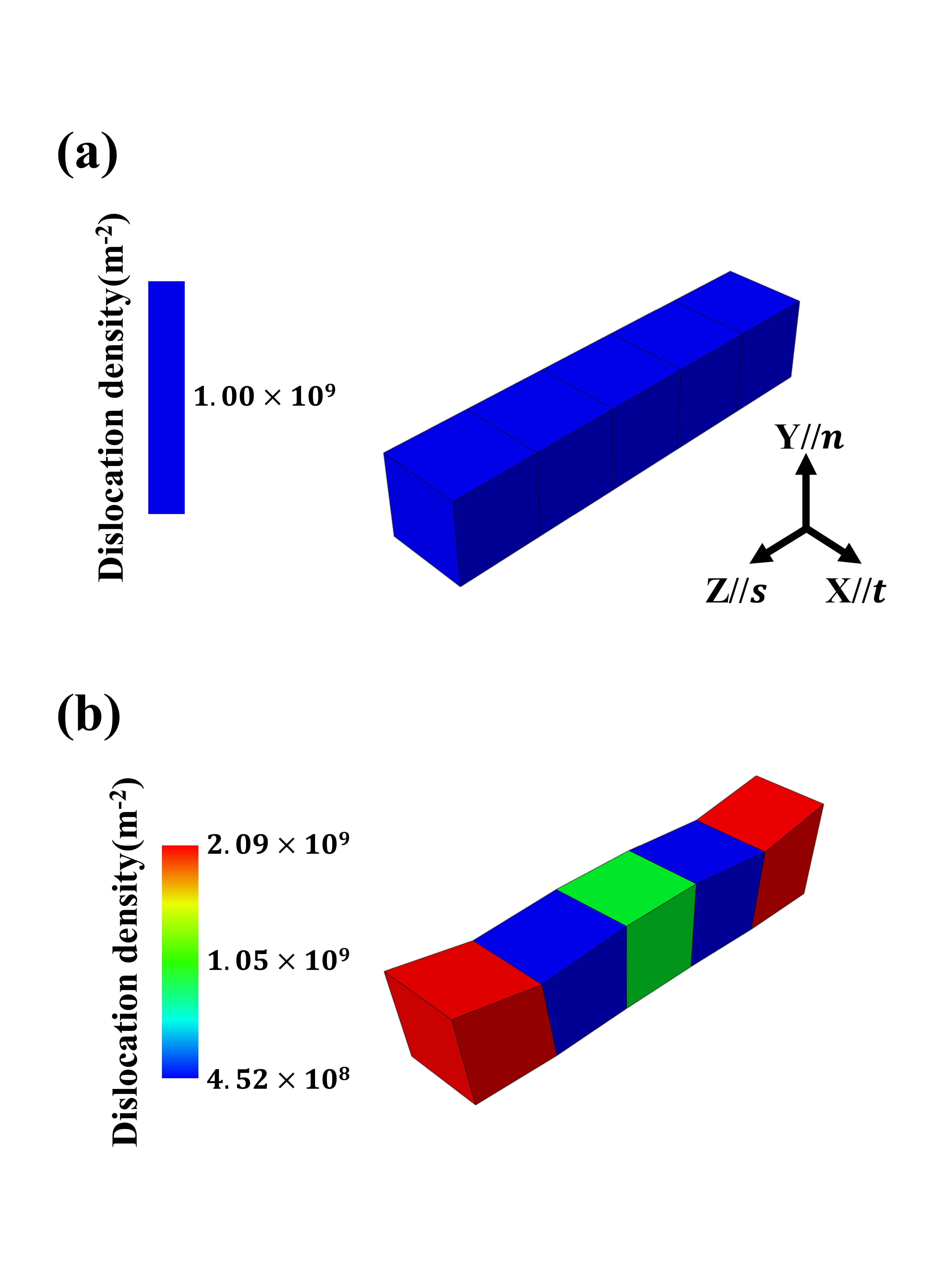}
\caption{Simulation results from original SD model~\cite{LIM20111328} considering only a single slip system with edge dislocations. (a) Initial state: dislocations are uniformly distributed without external loading. (b) After relaxation: due to mutual repulsion, dislocations migrate toward the ends of the domain, resulting in localized slip deformation near the boundaries. The coordinate arrows indicate directions: $\textbf{s}$ is the slip direction (Burgers vector direction), $\textbf{n}$ is the slip plane normal, and $\textbf{t}$ is the transverse direction.}\label{fig:five_elems}
\end{figure}

Lastly, \cref{fig:five_elems} shows the simulation results from the SD model considering only a single slip system with edge dislocations. The initial state exhibits uniformly distributed dislocations without external loading. After relaxation, dislocations migrate toward the ends of the domain due to mutual repulsion, resulting in localized slip deformation near the boundaries. These results indicate that additional work is required to find the equilibrium state of dislocations in the RVE, which was not considered in the original model~\cite{LIM20111328}. If the equilibrium state is determined, the PK force generated by the additional movement of dislocations can be interpreted as the PK force caused by the distortion of the elastic domain, which aligns with the DCM approach. Consequently, explicit PK force calculations are unnecessary, as the PK force can be implicitly included in the CPFEM framework.

To summarize, CPFEM with consistent continuum coupling already incorporates elastic dislocation interactions via internal plastic strain fields. Explicit PK force superposition is not only redundant but also destabilizing and inefficient. However, the SD model offers one key insight: each integration point in CPFEM effectively represents a population of dislocations. Therefore, intrapoint dislocation interactions should be modeled, and the back stress $\tau_{\text{back}}^a$ in \cref{eq:slip_rate} should arise from this internal structure. A refined model for this is introduced in \cref{GND induced back stress}.

\subsection{RVE and PBC}
\label{RVE and PBC}

\begin{figure}[!htbp]
\centering
\includegraphics[width=\textwidth]{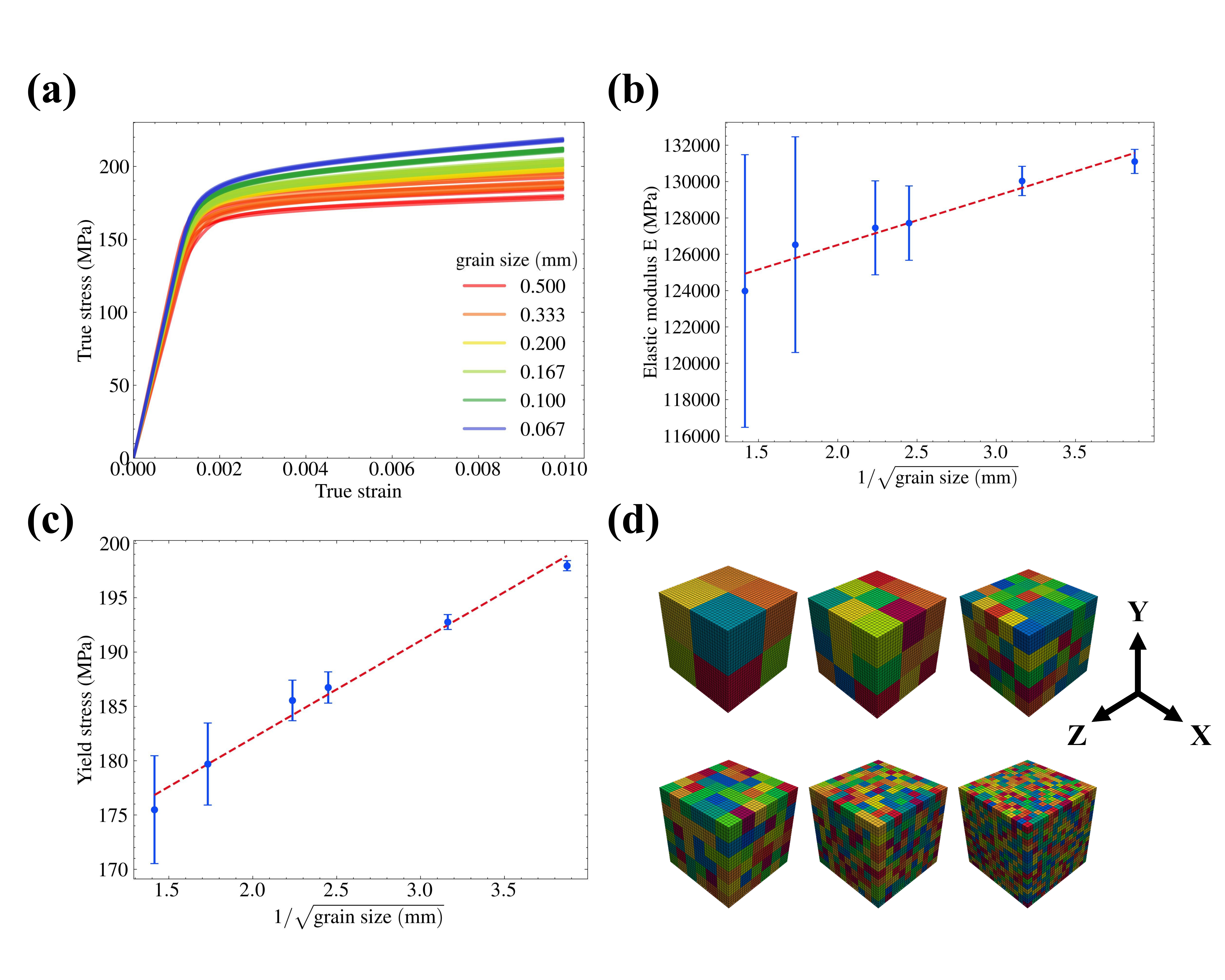}
\caption{
Mesh-induced strengthening artifacts.
(a) Flow stress increases as the number of grains increases as grain size decreases.
(b) The proportion of elements adjacent to grain boundaries rises, amplifying the influence of element stiffness and increasing overall elastic stiffness.
(c) Both the flow stress and yield stress increase, resulting in an artificial Hall-Petch-like relationship.
(d) All six cases were repeated 10 times with randomly assigned grain orientations. Each edge of the RVE is 1~mm. PBC was applied to all external faces of the RVE. SCCE-D model were used with the same material parameters (adopted from \cite{MIN2020102644}).
}
\label{fig:mesh-induced_strengthening_artifacts}
\end{figure}

RVE is a small, finite volume of material that statistically represents the overall behavior of a larger heterogeneous medium. In CPFEM, RVEs are constructed to capture key microstructural features such as grain size, shape, and crystallographic orientation. Each grain within the RVE is assigned a crystal orientation according to a target Orientation Distribution Function (ODF), which can be experimentally obtained using Electron Backscatter Diffraction (EBSD) or synthetically defined, as illustrated in \cref{fig:mesh-induced_strengthening_artifacts}(d).

PBCs are often applied to the external faces of the RVE to ensure that the material response is representative of an infinite medium. PBCs allow for the simulation of bulk material behavior by replicating the microstructural features across the boundaries, effectively eliminating edge effects and ensuring that the stress and strain fields are consistent across the RVE. All RVE simulations in this study were performed with PBCs, using multi-point constraints and formulations provided in Tian et al.~\cite{TIAN20191}.

The Hall-Petch relationship describes how the yield stress of a material increases with decreasing grain size:

\begin{equation}
\label{eq:Hall-Petch}
\sigma_y = \sigma_0 + k_d d^{-1/2},
\end{equation}

where \(\sigma_y\) is the yield stress, \(\sigma_0\) is the friction stress, \(k_d\) is the Hall-Petch slope, and \(d\) is the average grain size.

In CPFE simulations, an artificial Hall-Petch-like behavior may emerge even when the CPFEM does not include any size dependency, as in the SCCE-D model~\cite{LEE2010925}. This is due to mesh-induced artifacts, where the proportion of elements adjacent to grain boundaries (with perfect bonding) increases as the grain size decreases. As a result, the elastic stiffness of the RVE increases, leading to an increase in flow stress and yield stress, as shown in \cref{fig:mesh-induced_strengthening_artifacts}(b) and (c).

To demonstrate this, each RVE configuration in \cref{fig:mesh-induced_strengthening_artifacts}(d) was simulated ten times with randomly generated grain orientations. The collective results in \cref{fig:mesh-induced_strengthening_artifacts}(a) clearly show an increasing trend in flow stress with the number of grains. As shown in \cref{fig:mesh-induced_strengthening_artifacts}, the flow stress increases with decreasing grain size, despite identical material parameters (adopted from \cite{MIN2020102644}). This behavior is not intrinsic to the CPFEM model but instead results from mesh-induced artifacts: a larger proportion of stiff boundary-adjacent elements disproportionately influences the global response.

Lim et al.~\cite{LIM2019101} demonstrated that to achieve mesh-independent CPFEM results, an RVE should contain at least $10^4$ elements per grain and over $10^3$ grains in total. However, such requirements are computationally prohibitive for parameter calibration based on experimental data.

In this study, to strike a balance between accuracy and computational efficiency, we set the element size to one-tenth of the average grain size, ensuring a consistent number of elements per grain across all microstructures. This strategy effectively suppresses mesh-induced artifacts while maintaining tractable simulation costs. This approach is further demonstrated in \cref{EBSD results and RVEs}.

\section{GND-based back stress model and upscaled HAH model}
\label{GND-based back stress model}

In this section, we introduce a GND-based back stress model that captures the effects of dislocation wall structures and slip system orientation. The model, as discussed in \cref{introduction}, accounts for dislocation interactions within a single integration point, rather than between integration points. It incorporates dislocation wall structures and GND density, making the tensile mechanical response grain size-dependent. This explains the different stress-strain curves of microstructures with distinct grain sizes, ultimately leading to a unique parameter set determined by two samples: as-received and annealed. This is a key aspect of our methodology. Additionally, we present the HAH model, an anisotropic hardening model with Bauschinger effect and permanent softening, to upscale the GND-based back stress for comparing CPFEM-predicted T-C behavior with T-B experimental results.

\subsection{Dislocation density components and evolution}
\label{Dislocation density components and evolution}

\begin{figure}[!htbp]%% placement specifier
\centering
\includegraphics[width=0.8\textwidth]{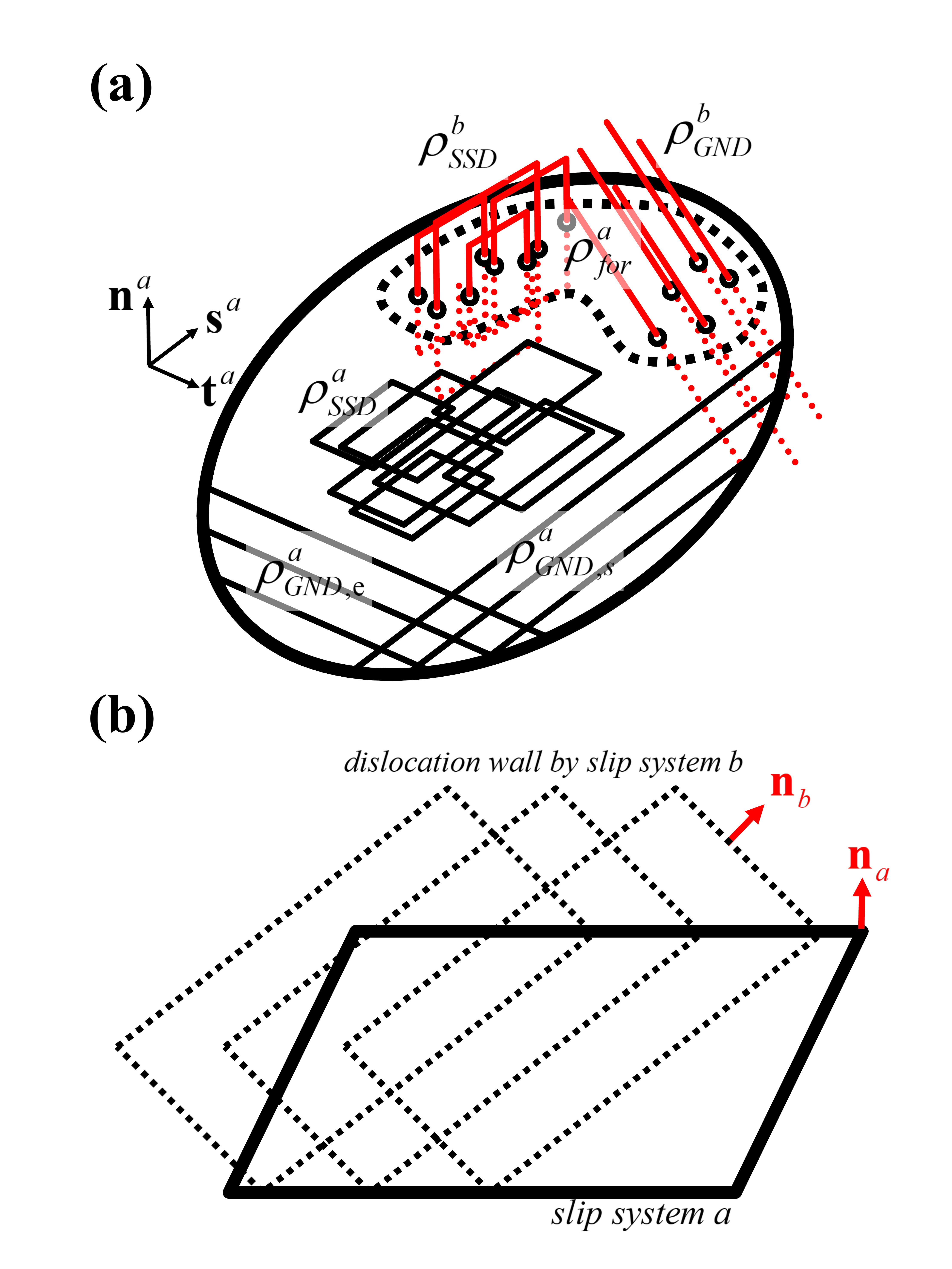}
\caption{Schematic illustration of dislocation density components within an integration point domain on slip system a. Dislocations are classified as edge or screw, and further into GND or SSD based on the net Burgers vector. Forest dislocation density arises from other slip systems $b$, weighted by the geometric alignment between their line directions and the slip plane normal of system a.}\label{fig:model_explanation}
\end{figure}
 
To clarify the concept of dislocation density in our model, \cref{fig:model_explanation}(a) illustrates the domain associated with a virtual slip system $a$, representing the region occupied by a single integration point in space. Within this domain, dislocation density can be classified into two categories: Statistically Stored Dislocation (SSD) density and GND density. These categories are distinguished based on the net Burgers vector of dislocation segments. In this framework, dislocations are assumed to be either pure edge or pure screw, with no mixed character. Based on this assumption, when the Burgers vectors of all edge or screw dislocation segments are summed, the net (non-zero) Burgers vector in each character is attributed to the GND density. The remaining dislocation segments, whose Burgers vectors cancel out, are accounted for as SSD.

The forest dislocation density on slip system $a$ arises from dislocations on other slip systems $b$. To account for their contribution, the model incorporates the geometric alignment between active slip system $a$ and the dislocations on system $b$, using the cosine of the angle between the slip plane normal of system $a$ and the line directions of edge and screw dislocations on system $b$. The expression for the forest dislocation density is given by:

\begin{equation}
\label{eq:rho_forest}
\rho _{{\rm f}}^{a} =\sum _{b}\left(\left|\boldsymbol{\mathrm{n}}^{a} \cdot \boldsymbol{\mathrm{t}}^{b} \right|\rho _{{\rm t,e}}^{b} +\left|\boldsymbol{\mathrm{n}}^{a} \cdot \boldsymbol{\mathrm{s}}^{b} \right|\rho _{{\rm t,s}}^{b} \right) ,
\end{equation}

where the total edge ($\rho _{{\rm t,e}}^{b}$) and screw ($\rho _{{\rm t,s}}^{b}$) dislocation densities on slip system $b$ are defined as:

\begin{equation}
\label{eq:rho_total}
\begin{array}{ll} {\rho _{{\rm t,e}}^{b} } & {=\rho _{{\rm SSD,e}}^{b} +\rho _{{\rm GND,e}}^{b} }, \\ {\rho _{{\rm t,s}}^{b} } & {=\rho _{{\rm SSD,s}}^{b} +\rho _{{\rm GND,s}}^{b} }. \end{array}
\end{equation}

Here, \textbf{$\boldsymbol{\mathrm{n}}^{b} $} denotes the slip plane normal of system $b$, and \textbf{$\boldsymbol{\mathrm{t}}^{b} $} represents the line directions of edge and screw dislocations on system $b$, respectively. This formulation captures the influence of the spatial orientation between interacting slip systems on the forest dislocation density. Assuming that the SSD dislocations are distributed in a square network, the initial edge and screw dislocation densities are taken to be equal, such that:

\begin{equation}
\label{eq:rho_ssd}
\textbf{$\left(\rho _{{\rm SSD,e}}^{b} \right)_{0} =\left(\rho _{{\rm SSD,s}}^{b} \right)_{0} =\frac{1}{2} \left(\rho _{{\rm SSD}}^{b} \right)_{0} $}.
\end{equation}

Based on the concept of forest dislocation density, the average spacing between dislocations that hinder slip on slip system $a$ is assumed to be inversely proportional to the square root of the forest dislocation density \(\rho _{{\rm f}}^{a}\). Since slip resistance is inversely related to this spacing at the mesoscale, the critical resolved shear stress on slip system $a$ can be expressed following the Taylor hardening law.

\begin{equation}
\label{eq:g}
g^{a} =\alpha Gb \sqrt{\rho _{{\rm f}}^{a} },
\end{equation}

where $\alpha$ is the geometric factor, $G$ is the shear modulus, and $b$ is the magnitude of the Burgers vector.

The evolution of dislocation density is statistically described by the KM evolution law~\cite{KOCKS2003171}. In this study, GND density is not treated as arising independently but is instead considered a part of the total dislocation density generated during plastic deformation by the KM law. In other words, a portion of the increasing total dislocation density is converted into GNDs, while the decrease in GND density also contributes to an increase in SSD. Based on this concept, the evolution of dislocation density is modeled in two steps: the total dislocation density is updated according to the KM law, as in \cref{eq:KM1}, while the partitioning between GND and SSD is determined based on the change in incompatibility, as in \cref{eq:KM2}.

\begin{equation}
\label{eq:KM1}
\dot \rho _{KM}^{a} =\left(k_{1} \frac{\sqrt{\rho _{{\rm f}}^{a} } }{b^{a} } -k_{2} \rho _{{\rm SSD}}^{a} \right)|\dot \gamma ^{a} |,
\end{equation}

\begin{equation}
\label{eq:KM2}
\begin{array}{l} {\dot \rho _{SSD,e}^{a} =\frac{1}{2} \dot \rho _{KM}^{a} -\dot \rho _{GND,e}^{a} }, \quad {\dot \rho _{SSD,s}^{a} =\frac{1}{2} \dot \rho _{KM}^{a} -\dot \rho _{GND,s}^{a} }, \end{array}
\end{equation}

where $\dot \rho _{KM}^{a}$ is the total dislocation generation rate on slip system \(a\), and $\dot \rho _{GND,e}^{a}$, $\dot \rho _{GND,s}^{a}$ denote the increase in GND formation in edge and screw components, respectively. The remaining portion is assigned to SSD evolution under the assumption of equal partitioning between edge and screw characters. Here, $k_{1}$ and $k_{2}$ are material constants.

\subsection{Calculation of GND density in finite element framework}
\label{GND calculation method}

The GND densities $\rho _{{\rm GND},{\rm e}}^{a} $ and $\rho _{{\rm GND},{\rm s}}^{a} $ at each time increment are computed using the total form of incompatibility based on the curl of the plastic deformation gradient, as described by Model 1-a in \cite{DEMIR2024104013}. The incompatibility tensor $\boldsymbol{\mathrm{\Lambda }}$ is given by

\begin{equation}\label{eqn:Nye1}
\boldsymbol{\mathrm{\Lambda }}=-(\nabla \times \boldsymbol{\mathrm{F}}_{p} )^{T} ,
\end{equation}

which must be fully accommodated by the Nye tensor associated with GNDs as follows:

\begin{equation}
\boldsymbol{\mathrm{\Lambda }}=\sum _{a}\rho _{{\rm GND}}^{a}  b^{a} \boldsymbol{\mathrm{s}}^{a} \otimes \boldsymbol{\mathrm{l}}^{a},
\end{equation}

where $\boldsymbol{\mathrm{l}}^{a} $ is the dislocation line vector on slip system $a$, which separately accounts for edge and screw systems. This tensorial equation is reformulated into a linear system:
\begin{equation}
\{ \boldsymbol{\mathrm{\Lambda }}\} =[\boldsymbol{\mathrm{A}}]\{ \rho _{{\rm GND}} \},
\end{equation}
with the coefficient matrix containing the dyadic geometry terms associated with each dislocation family. Since the number of dislocation types (e.g., edge and screw components for each slip system) typically exceeds the number of independent components in the incompatibility tensor, the resulting system of equations is underdetermined. This system is therefore solved using the Moore--Penrose pseudoinverse~\cite{MOORE1920, PENROSE1955}:
\begin{equation}
\{ \rho _{{\rm GND}} \} =[\boldsymbol{\mathrm{A}}]^{+} \{ \boldsymbol{\mathrm{\Lambda }}\} ,\quad {\rm where}\quad [\boldsymbol{\mathrm{A}}]^{+} =[\boldsymbol{\mathrm{A}}]^{T} ([\boldsymbol{\mathrm{A}}][\boldsymbol{\mathrm{A}}]^{T} )^{-1} .
\end{equation}
This corresponds to the $L_2$-minimization approach, which yields the GND distribution of minimal line length satisfying the incompatibility constraint.

Accurate computation of the curl operator in \cref{eqn:Nye1} requires reliable evaluation of spatial gradients of the plastic deformation gradient field. This section describes the procedure used to compute the gradient operator in 8-node linear hexahedral (C3D8) elements, which are widely used in 3D FE simulations. For C3D8 elements in ABAQUS~\cite{abaqus2024}, the number of nodes $p = 8$ is the same as the number of integration points $q = 8$, satisfying the condition $p = q$. This enables a direct mapping from integration point values to nodal values via an identity transform, eliminating the need for extrapolation or pseudoinverse fitting.

The curl of the plastic deformation gradient $\nabla \times \boldsymbol{\mathrm{F}}^{p}$ is computed at each integration point through the following steps:
\begin{itemize}

\item   The plastic deformation gradient $\boldsymbol{\mathrm{F}}^{p}$ is evaluated at each integration point within the element as part of the stress update procedure in the user-defined material subroutine, UMAT.

\item   At each integration point, the gradient operator $[\boldsymbol{\mathrm{T}}]_{3\times p}$ is computed using the derivatives of shape functions with respect to the isoparametric coordinates:
\begin{equation}
[\boldsymbol{\mathrm{T}}]_{3\times p}=[\boldsymbol{\mathrm{J}}^{-T}]_{3\times 3} [\nabla \boldsymbol{\mathrm{N}}]_{3\times p},
\label{eqn:eqT}
\end{equation}
where the Jacobian matrix $[\boldsymbol{\mathrm{J}}]_{3\times 3}$ is given by: 
\begin{equation}
[\boldsymbol{\mathrm{J}}^{T}]_{3\times 3}=[\nabla \boldsymbol{\mathrm{N}}]_{3\times p}  [\boldsymbol{\mathrm{x}}]_{p \times 3},
\label{eqn:eqJ}
\end{equation}
where $[\nabla \boldsymbol{\mathrm{N}}]_{3\times p}$ contains the shape function derivatives in isoparametric space and $[\boldsymbol{\mathrm{x}}]_{p \times 3}$ are physical nodal coordinates.

\item  Since the number of nodes equals the number of integration points for C3D8 elements with $p = q = 8$, nodal values can be directly obtained from integration point data via
\begin{equation}
[\boldsymbol{\mathrm{M}}]_{p \times q}=[\boldsymbol{\mathrm{N}}^{-1}]_{p \times q},
\end{equation}
where $[\boldsymbol{\mathrm{N}}]_{p\times q}$ is the matrix for shape functions evaluated at integration points. Then, the gradient tensor $\boldsymbol{\mathrm{G}}$ is calculated as follows:
\begin{equation}
[\boldsymbol{\mathrm{G}}]_{3\times q} =[\boldsymbol{\mathrm{T}}]_{3\times p} [\boldsymbol{\mathrm{M}}]_{p\times q}.
\end{equation}
\end{itemize} 

It is noted that the value of $[\boldsymbol{\mathrm{T}}]_{3\times p}$ is different at every integration point, since the isoparametric coordinates used in \cref{eqn:eqT} and \cref{eqn:eqJ} are different. 
When the $n$th integration point's $[\boldsymbol{\mathrm{T}}]_{3\times p}$ is $[\boldsymbol{\mathrm{T}}]_{3\times p}^n$ and the corresponding $[\boldsymbol{\mathrm{G}}]_{3\times q}$ is $[\boldsymbol{\mathrm{G}}]_{3\times q}^n$, then the spatial derivative of $\boldsymbol{\mathrm{F}}_{p}$ at each integration point is
\begin{equation}
\left[\begin{array}{c}
(F_{ij}^p )_{,1}^{n} \\
(F_{ij}^p )_{,1}^{n} \\
(F_{ij}^p )_{,2}^{n} \\
(F_{ij}^p )_{,3}^{n} 
\end{array}\right]
=
[\boldsymbol{\mathrm{G}}]_{(3\times q)}^{n}
\left[\begin{array}{c}
(F_{ij}^p )^{1} \\
(F_{ij}^p )^{2} \\
\vdots \\
(F_{ij}^p )^{q}
\end{array}\right].
\end{equation}
Finally, the curl of the plastic deformation gradient $(\nabla \times \boldsymbol{\mathrm{F}}^{p})^n $ is evaluated at each integration point by the standard definition for a curl of second-order tensor: $(\nabla \times \mathrm{F}^{p} )_{ij}^n =\varepsilon _{irs} (\mathrm{F}_{sj}^{p} )^n_{,r} ,$  where $\varepsilon _{irs} $ is the Levi-Civita permutation symbol.

\subsection{GND induced back stress}
\label{GND induced back stress}

Geometrically Necessary Boundaries (GNBs) are incorporated into the model as the physical origin of back stress. Under non-uniform plastic deformation, as shown in \cref{GND calculation method} (see also \cite{DEMIR2024104013}), an excess density of same-signed dislocations (GNDs) is required to maintain lattice continuity. These GNDs tend to organize into planar arrays separating regions with different slip patterns, forming GNBs that accommodate the local lattice misorientation~\cite{HUGHES2003147,Yu2020}. In contrast to incidental dislocation boundaries (IDBs) that form by random trapping of statistically stored dislocations, GNBs carry a net Burgers vector (due to the GND content) commensurate with the lattice rotation between adjacent regions~\cite{HUGHES2003147}. The TEM study by Hughes and Hansen confirms that dislocations assemble into GNBs with significant misorientations at large strain, dispelling earlier notions that internal walls are dipolar structures with no net Burgers vector~\cite{HUGHES2003147}. Because GNBs are enriched in GNDs, they act as robust internal barriers to slip. That is, dislocations piling up against a GNB on one side create a long-range internal stress field that extends across the boundary~\cite{KASSNER201344}. This was also experimentally validated by Hasegawa et al.~\cite{HASEGAWA1975267}, who showed via TEM that dislocation structures developed during pre-straining—such as aligned walls and cells—impede reverse slip and thereby induce internal back stresses. Their study demonstrated that the Bauschinger effect becomes more pronounced as these internal boundaries accumulate, confirming the critical role of internal dislocation arrangements, including GNB-like features, in controlling kinematic hardening during strain path reversal.

In the literature, several models have been proposed to capture the cell walls or GNBs. Aoyagi et al.~\cite{AOYAGI201313} proposed a CPFEM model combined with a reaction-diffusion model on dislocation patterning to capture the formation of dislocation cell structures. Viatkina et al.~\cite{VIATKINA2007982, VIATKINA20076030} treated the cell and wall as a composite, with the back stress arising from the interaction between the cell and wall dislocations. Although these models provided insights into dislocation structures and their effects on the stress-strain response, they are not suitable for our goal of identifying a unique parameter set using only two microstructures, as they require a large number of parameters to be calibrated. Instead, we treat GNBs in the same manner as grain boundaries, which allows the application of the analytical pile-up polarization model proposed by Liu et al.~\cite{LIU2022104793}. In their study, the back stress arising from asymmetric resistance across boundaries is modeled as follows:

\begin{equation}
\label{eq:back stress_LIU2022104793}
\tau_{back}^a = \frac{ G h \min(\gamma^a, \gamma^{\text{sat}}) }{ \pi (1-\nu) D} \text{sgn}(\gamma^a),
\end{equation}

where $h$ is the average distance between activated slip systems and $D$ is the average grain size. In this context, the average grain size $D$ is replaced with the average GNB spacing. The spacing is inversely proportional to the square root of the GND density, similar to the Taylor hardening law. However, the density over all slip systems should be considered carefully since the angles between all slip systems are different. This relationship is explained in \cref{fig:model_explanation}(b), where each slip system $b$ contributes to the GND density on slip system $a$ based on the geometric alignment between their slip plane normals and line directions. The proposed back stress is then expressed as:

\begin{align}
\label{eq:back stress}
\tau_{back}^a &= \frac{ \zeta \min(\gamma^a, \gamma^{\text{sat}}) }{ \gamma^{\text{sat}} } G b \notag \\ 
&\quad \cdot\left[ \sum_b \left( (\mathbf{n}_a \cdot \mathbf{n}_b) \left( \frac{ \sqrt{ |\rho^{b}_{\text{GND,e}} |} }{ 1 - \nu } + \sqrt{|\rho^{b}_{\text{GND,s}}|} \right) \right) \right] \text{sgn}(\gamma^a).
\end{align}

The above model introduces two parameters: the saturation threshold $\gamma^{\text{sat}}$ and the backstress parameter $\zeta$. Since GNBs exhibit relatively low resistance and their spacing is narrower than typical grain boundaries, the polarization can saturate quickly. In this study, the value was assumed to be $0.04$ based on preliminary numerical tests varying its value from $0.01$ to $0.08$, which resulted in negligible effect on the overall stress-strain response, as shown in \cref{minor_parameters}. Therefore, the only parameter requiring calibration is $\zeta$, which governs the magnitude of the back stress. \Cref{eq:back stress} becomes the same form as \cref{eq:back stress_KAPOOR2018447} when $\gamma^a$ exceeds $\gamma^{\text{sat}}$.

This parameter can be identified using uniaxial tensile stress-strain data from two microstructures of the same material but with different grain sizes. Given that GND density scales inversely with grain size, the difference in stress-strain response provides sufficient information to optimize $\zeta$. Once calibrated, the model enables predictive simulations of tension-compression behavior and strain-path sensitivity in ultra-thin polycrystalline sheets, using only tensile test data.

\subsection{Upscaled HAH model with permanent softening}
\label{Upscaled HAH}

While the GND density-based CPFEM model with RVEs is advantageous in capturing detailed microstructural behavior, it inherently requires a polycrystalline representation where each grain is discretized by multiple finite elements. This ensures that local strain gradients and GND densities are accurately resolved within each grain. Unlike homogenized schemes such as the isostrain (Taylor) or isostress (Sachs) assumptions, or mean-field models like VPSC~\cite{LEBENSOHN19932611}, which allow multiple grain orientations per integration point, our approach mandates that each grain is represented by many elements. This is necessary because the evaluation of GND density relies on spatial gradients of the plastic deformation field at the grain scale, making such approximations inapplicable. In other words, a fully resolved polycrystalline RVE with many elements per grain is essential for our model. To accurately simulate tension-bending at the millimeter scale, an RVE of comparable size would be required. For instance, modeling a $1\,\text{mm}^3$ volume with an average grain size of $10\,\mu\text{m}$ would necessitate approximately $10^6$ grains. This level of resolution is computationally prohibitive for finite element analysis.

To efficiently upscale the tension-compression responses obtained from CPFEM simulations to the macroscale, we employed the HAH model, originally proposed by Barlat et al.~\cite{BARLAT2020947}, and further developed to include cross-loading effects, permanent softening, or to reduce the number of parameters~\cite{BARLAT2013130, BARLAT2014201, REYNE2022103303}. The HAH model captures path-dependent phenomena such as stress overshoot, cross-loading contraction, and directional hardening by allowing the yield surface to distort while preserving a homogeneous functional form. This provides a physically meaningful and computationally efficient framework for simulating non-proportional loading at the macroscale. The HAH model was calibrated using the upscaled tension-compression responses from CPFEM, as described in \cref{T-C test}.

Furthermore, to reflect the permanent softening behavior observed in our CPFEM simulations (see \cref{shear_reverse_shear}), we used an extended version of the original HAH model by introducing a scalar degradation variable $g_p$, which reduces the flow strength as a function of accumulated plastic strain. This modification enables the model to capture both transient hardening during load path changes and irreversible softening under cyclic loading, while preserving the geometric consistency of the evolving yield surface.

\begin{figure}[!htbp]
     \centering
     \includegraphics[width=\textwidth]{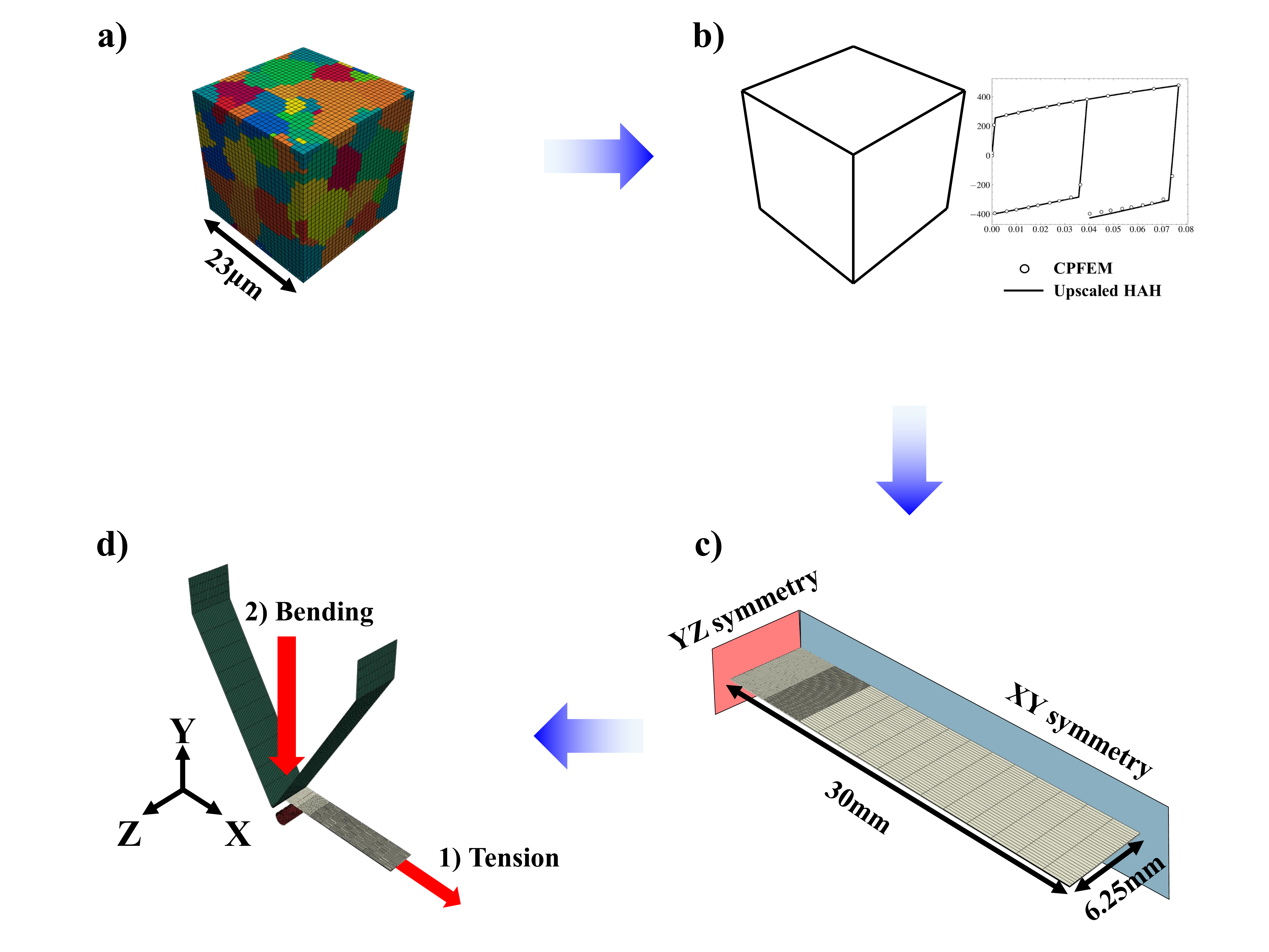}
     \caption{Overview of the upscaling procedure for tension-bending simulation. (a) CPFEM RVE simulation captures grain-scale GND evolution and tension-compression response . (b) The upscaled HAH model is calibrated to reproduce the CPFEM macroscopic stress-strain behavior. (c) Guage section of the experimental tension-bending specimen in FEM (d) Finite element simulation of the tension-bending test using the upscaled HAH model, enabling direct comparison with experimental force-displacement data.}
     \label{fig:upscale_overview}
\end{figure}
The overall upscaling workflow is illustrated in \cref{fig:upscale_overview}. 
First, CPFEM RVE simulations are performed to capture the grain-scale evolution of GND density and the corresponding tension-compression response. 
Next, the HAH model parameters are calibrated to reproduce the macroscopic stress-strain behavior observed in CPFEM.
For the FE simulation of the tension-bending test, only the gauge section of the specimen is modeled, with the thickness set identical to the actual specimen. 
The mesh consists of 10 elements through the thickness direction, and a total of 83,100 C3D8R elements of ABAQUS~\cite{abaqus2024} are used.
In the simulation procedure, tension is applied to the specimen (0\%, 4\%, and 8\% engineering strain), followed by unloading and bending.
The upscaled HAH model is used in FE simulations of the tension-bending test geometry, enabling direct comparison with experimental force-displacement curves and validation of the predictive capability of the multiscale modeling approach.

The yield function of the HAH model is defined as:

\begin{align} 
\Phi(\boldsymbol{s}) \ & =  \frac{\left[ \phi^q + \phi_h^q \right]^{1/q}}{g_p} \notag \\ 
& = \frac{\left\{ \left[ \phi(\boldsymbol{s}) \right]^q + f_1^q \left| \hat{\boldsymbol{h}}^s : \boldsymbol{s} - \left| \hat{\boldsymbol{h}}^s : \boldsymbol{s} \right| \right|^q + f_2^q \left| \hat{\boldsymbol{h}}^s : \boldsymbol{s} + \left| \hat{\boldsymbol{h}}^s : \boldsymbol{s} \right| \right|^q \right\}^{1/q}}{g_p} \notag \\
& = \sigma(\bar{\varepsilon}),
\end{align}

where \( \boldsymbol{s} \) is the deviatoric part of the Cauchy stress tensor, \( \phi(\boldsymbol{s}) \) is the isotropic component (in this case, the von Mises yield function~\cite{vonMises1913}), \( \hat{\boldsymbol{h}}^s \) is the microstructure deviator, \( \Phi_h(\boldsymbol{s}) \) is a directional hardening term, \( g_p \) is the scalar softening parameter, \( \bar{\sigma} \) is the equivalent stress, and \( q \) is a material parameter. For simplicity, we set \( f_1 = f_2 = f \).

The evolution of microstructure deviator with respect to the equivalent plastic strain \( \bar{\varepsilon}^p \) is given by:
\begin{equation}
\frac{d \hat{\boldsymbol{h}}^s}{d \bar{\varepsilon}^p} 
= k \left[ -\hat{\boldsymbol{s}} + \frac{8}{3} \hat{\boldsymbol{h}}^s (\hat{\boldsymbol{s}} : \hat{\boldsymbol{h}}^s) \right] 
\text{sgn}(\boldsymbol{s} : \hat{\boldsymbol{h}}^s),
\end{equation}
where the normalized deviatoric stress is \( \hat{\boldsymbol{s}} = \boldsymbol{s} / \sqrt{ \boldsymbol{s} : \boldsymbol{s} } \), and \( k \) is a material parameter. This evolution law ensures that \( \hat{\boldsymbol{h}}^s \) gradually aligns with \( \hat{\boldsymbol{s}} \), converging to \( \hat{\boldsymbol{s}} / \sqrt{8/3} \) in the limit of large plastic strain under constant loading.

The scalar function \( f \), dependent on the internal variable \( g \), modulates the asymmetry of the yield surface and is formulated as
\begin{equation}
f = \left( g^{-q} - 1 \right)^{1/q},
\end{equation}
and the evolution of \( g \) is governed by
\begin{equation}
\frac{dg}{d \bar{\varepsilon}^p} = k_g  \frac{\tau_g - g}{g^{\gamma_g}},
\end{equation}
where \( k_g \), \( \tau_g \), and \( \gamma_g \) are material parameters. The variable \( g \) controls the distortion of the yield surface during load reversals.

\begin{figure}[!htbp]
\centering
\includegraphics[width=\textwidth]{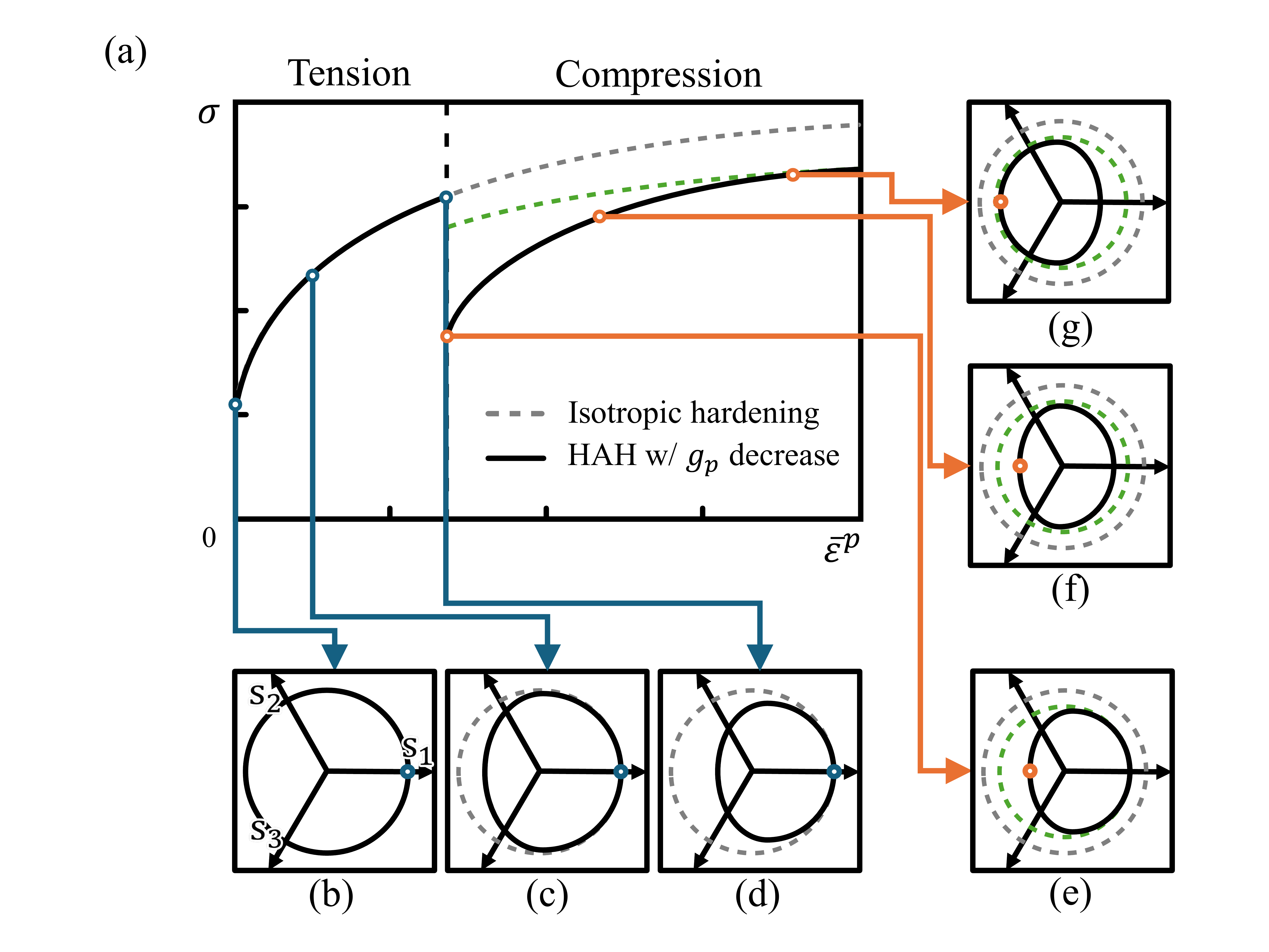}
\caption{Yield surface evolution of the HAH model with permanent softening under tension-compression loading. (a) Comparison of isotropic hardening and HAH model yield surfaces. (b)-(g) Sequential evolution of the yield surface, highlighting the effect of permanent softening by green dotted line.}
\label{fig:HAH_schematic}
\end{figure}

To model permanent softening, the evolution of \( g_p \) is expressed as
\begin{equation}
\label{eq:gp_evolution}
\frac{dg_p}{d \bar{\varepsilon}^p} = 
\max \left( 
    k_p (\tau_p - g_p) \left\langle -d\hat{\boldsymbol{\varepsilon}}^p : \hat{\boldsymbol{\varepsilon}}^p \right\rangle,
    - \frac{ \partial \sigma_R }{ \partial \bar{\varepsilon}^p } \cdot \frac{g_p}{\sigma_R}
\right),\;\; g_p(0) = 1,
\end{equation}
where \( \langle \cdot \rangle \) denotes the Macaulay bracket, and \( \sigma_R(\bar{\varepsilon}^p) \) is the isotropic reference hardening curve. The normalized plastic strain is defined as \( \hat{\boldsymbol{\varepsilon}}^p = \boldsymbol{\varepsilon}^p / \sqrt{ \boldsymbol{\varepsilon}^p : \boldsymbol{\varepsilon}^p } \). The first term in max operator captures that the softening only occurs when load reversal happens by dislocation annihilation, while the second term ensures that the global hardening rate does not become negative. Further details on the permanent softening can be found in \cite{REYNE2022103303}. A schematic illustration of the yield surface described by the HAH model, including the directional evolution governed by \( g \) and the isotropic degradation controlled by \( g_p \), is shown in Figure~\ref{fig:HAH_schematic}.

\vspace{1em}
 
This formulation enables the HAH model to capture both transient hardening during load path changes and permanent stress degradation due to plastic strain accumulation. The model is calibrated using CPFEM simulations and applied to macroscopic simulations involving tension-bending deformation.

\section{Experiments}
\label{Experimental}

The experimental procedure consisted of two categories based on material selection and testing methodology. In the first category, ferritic low-carbon steel sheets with a thickness of 0.64 mm were subjected to annealing to achieve two distinct grain size distributions. Tensile tests were performed to obtain stress-strain curves, followed by tension-compression tests to characterize the material's behavior under cyclic loading. In the second category, SUS316 stainless steel sheets with a thickness of 0.083 mm were annealed and tested under uniaxial tension. Due to the sheet's extreme thinness, direct compression testing was not conducted. Instead, bending tests were performed to evaluate the mechanical response under loading and unloading, focusing on force-displacement behavior and springback effects.

\subsection{Annealing and tensile test}
\label{Annealing and tensile test}

Annealing was performed to increase the grain size of both low-carbon ferritic steel and austenitic stainless steel, which is SUS316. Specimens were sealed in vacuum glass tubes (vacuum < 100 mTorr) to prevent oxidation and heat-treated in a Lindberg/Blue M Moldatherm$^{\mathrm{TM}}$ box furnace. For low-carbon steel, annealing was conducted at 900$^\circ$C, and for SUS316 at 1100$^\circ$C, both with a heating rate of 50$^\circ$C/min, held for 4 hours, followed by air cooling to room temperature.

Prior to annealing, low-carbon steel specimens were pre-strained to 1\% at a strain rate of 10$^{-3}$ s$^{-1}$ to promote grain growth and alter the subsequent stress-strain response. After pre-straining, the specimens were unloaded and then subjected to the annealing process as described above.

Tensile tests were performed on both as-received and annealed specimens using an Instron 8801 hydraulic drive system. A strain rate of 10${}^{-3}$ s${}^{-1}$ was applied until fracture. To characterize phase, texture, and grain size distribution, the specimens were cut, mounted, and polished for EBSD analysis. The measurements were conducted using a Velocity Ultra / EDAX EBSD system over a $1500m \mathrm{\times} 1500m$ area with a step size of $1.60 m$.

The chemical compositions of the materials are summarized in \cref{tab:chemical_composition}.

\begin{table}[htbp]
     \label{tab:chemical_composition}
\centering
\caption{Chemical compositions (wt.\%) of low-carbon steel and SUS316.}
\begin{tabular}{lcccccccc}
\toprule
Material & C    & Si   & Mn   & P    & S    & Ni   & Cr   & Mo   \\
\midrule
Low-carbon steel & 0.0012 & 0.005 & 0.078 & 0.012 & 0.054 & -- & -- & -- \\
SUS316           & 0.030  & 1.0  & 2.0  & 0.045 & 0.030 & 12. & 16. & 2.5 \\
\bottomrule
\end{tabular}
\end{table}

\subsection{Tension-Compression (T-C) test}
\label{T-C test}

\begin{figure}[!htbp]
\centering
\includegraphics[width=\textwidth]{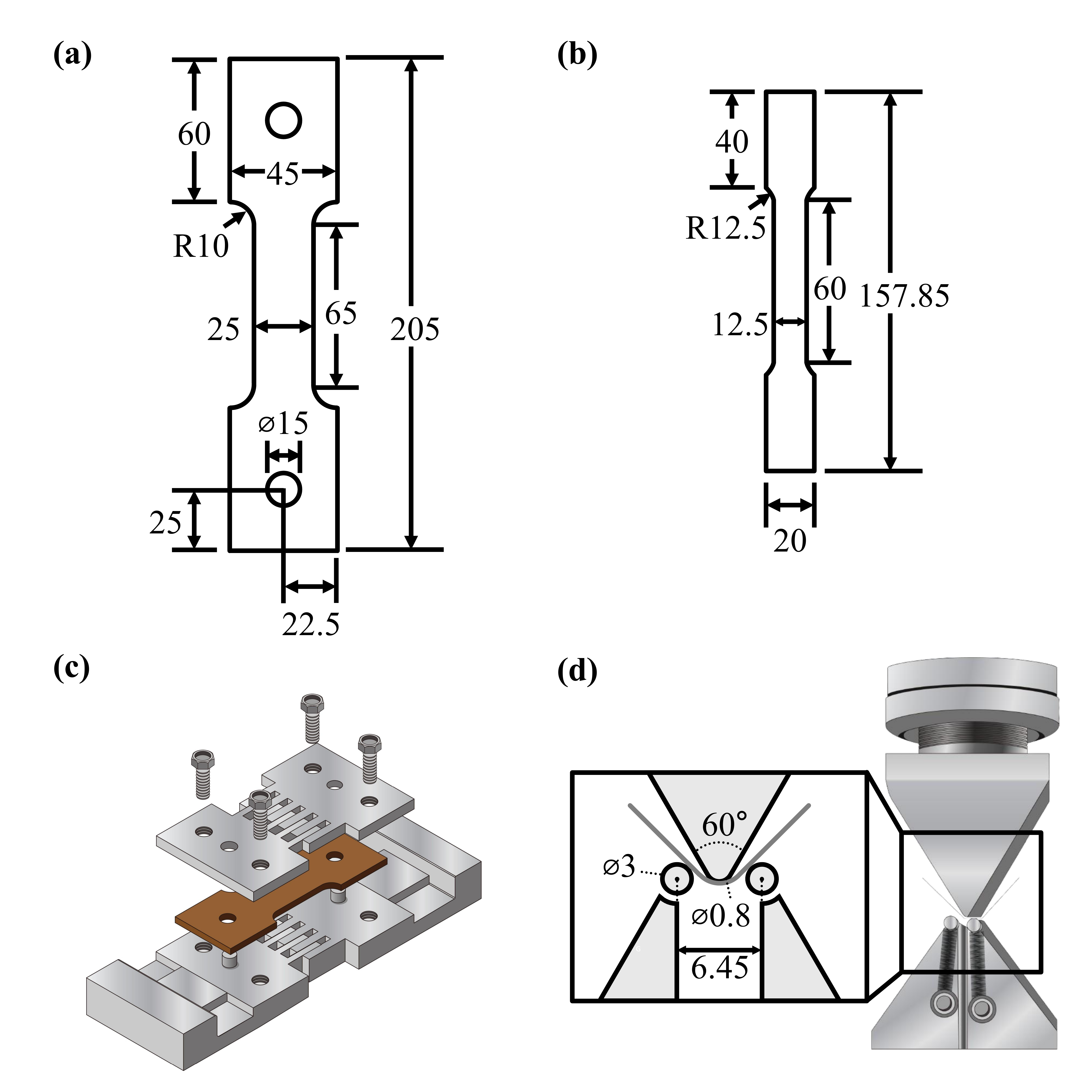}
\caption{
Schematics of experimental setups and specimen geometries.
(a) Dimensions of tension-compression test specimen.
(b) Dimensions of tension-bending test specimen.
(c) Assembly for tension-compression test with anti-buckling fixture.
(d) Three-point bending setup for tension-bending test. All length dimensions are in $mm$.
}
\label{fig:specimens}
\end{figure}

Tension-compression tests were performed on 0.64 mm-thick low-carbon steel using a testing apparatus developed by Boger et al.~\cite{BOGER20052319}. A specially designed tension-compression specimen (\Cref{fig:specimens}(a)) was machined from the as-received low-carbon steel. To prevent buckling during compression, a side force of 2 kN was applied using solid plates, with Teflon sheets and lubricant inserted between the specimen and plates to minimize friction.

The test was conducted at a strain rate of 10${}^{-3}$ s${}^{-}$${}^{1}$. The specimen was initially loaded in tension until an engineering strain of 0.1 was reached, after which it was reverse-loaded in compression until an engineering strain of 0.0. To account for frictional effects, the stress-strain curve from the tensile test was compared with the tensile segment of the tension-compression test, and the friction coefficient was calibrated to 0.04. Then, by the friction coefficient, tension-compression test's stress-strain curve was corrected. Details of the experimental setup are shown in \cref{fig:specimens}(a) and (c).

\subsection{Tension-Bending (T-B) test}
\label{T-B test}

The tension-bending test for evaluating kinematic or distortional hardening behavior in sheet metals was originally introduced by Zang et al.~\cite{ZANG201484}. The methodology was further used to obtain parameters of HAH model by Choi et al.~\cite{CHOI2019428}. Following this procedure, dog-bone-shaped specimens (\Cref{fig:specimens}(b)) were first pre-strained under uniaxial tension to engineering strains of 4\% and 8\%, along with an additional as-recieved specimen for comparison.

After pre-straining, the specimens were cut into rectangular samples for bending tests, ensuring consistent strain distribution across all cases, including the as-received specimen. The specimens were unloaded after reaching the target strain before being subjected to bending.

The prepared specimens were tested using a three-point bending setup, where a punch applied force at the center while the sample was supported at both ends. The punch stroke was fixed at 5 mm for all cases. The radius of the two supporting rollers was 1.5~mm, the punch tip radius was 0.4~mm, the punch angle was $60^\circ$, and the punch speed was 1~mm/min. Force-displacement curves obtained from the bending tests were analyzed to characterize the material's mechanical response under bending deformation. Details of the experimental setup are shown in \cref{fig:specimens}(b) and (d).

\section{Results}
\label{Results}

\subsection{EBSD results and RVEs}
\label{EBSD results and RVEs}

\begin{figure}[!htbp]
\centering
\includegraphics[width=\textwidth]{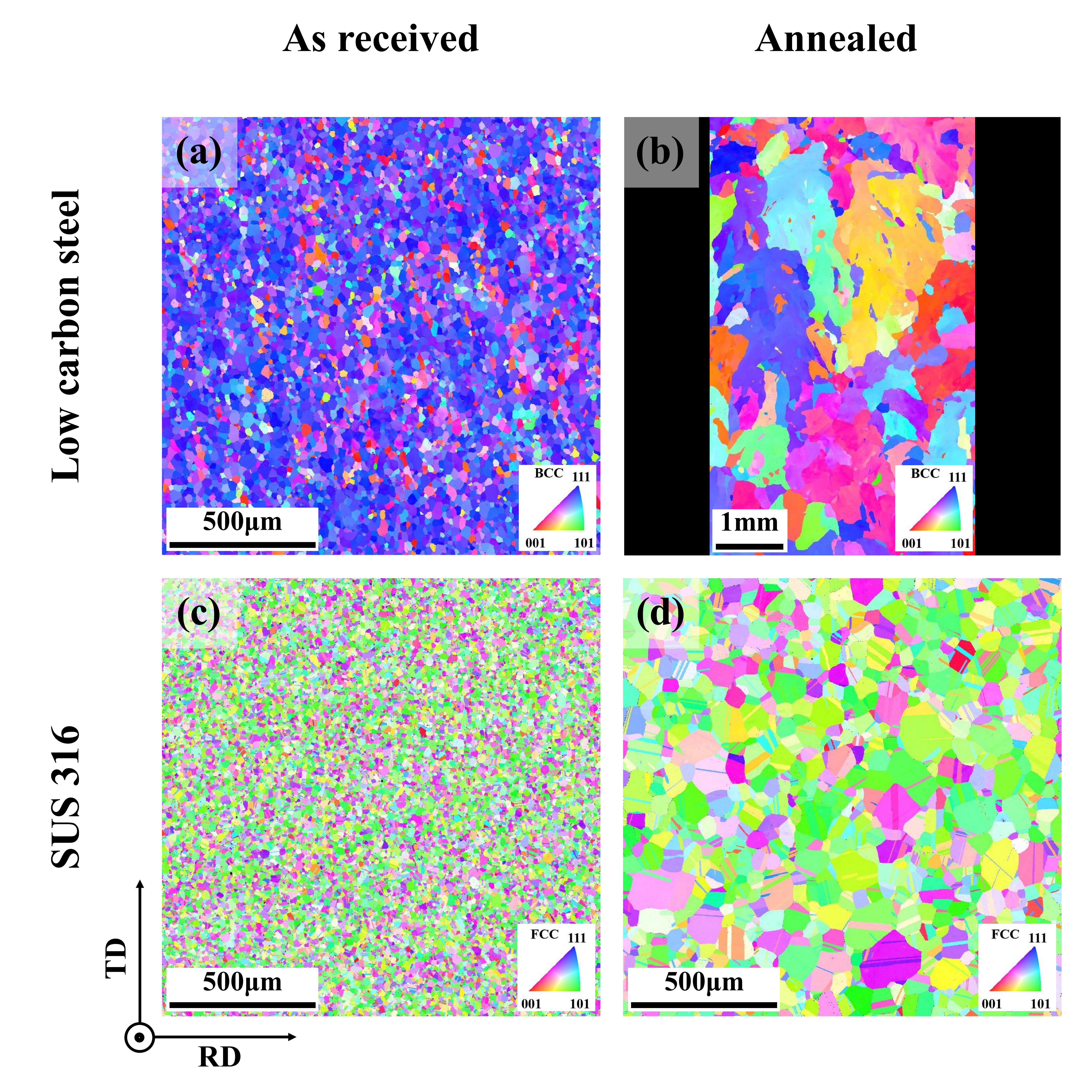}
\caption{Inverse pole figure map for (a) as-received low-carbon steel, (b) annealed low-carbon steel, (c) as-received SUS316, and (d) annealed SUS316. Annealing results in significant grain growth for both materials. The color coding represents crystallographic orientation according to the inverse pole figure (IPF) triangle shown in each panel. RD: Rolling Direction, TD: Transverse Direction.}\label{fig:EBSD}
\end{figure}
The EBSD maps in \cref{fig:EBSD} show the grain structures and crystallographic orientations for both low-carbon steel and SUS316 in as-received and annealed conditions. The annealing process resulted in significant grain growth for both materials, with average grain sizes of 42~$\mu$m for low-carbon steel and 23~$\mu$m for SUS316 in the as-received condition, increasing to around 330~$\mu$m and 78~$\mu$m, respectively, after annealing.

\begin{figure}[!htbp]
\centering
\includegraphics[width=0.6\textwidth]{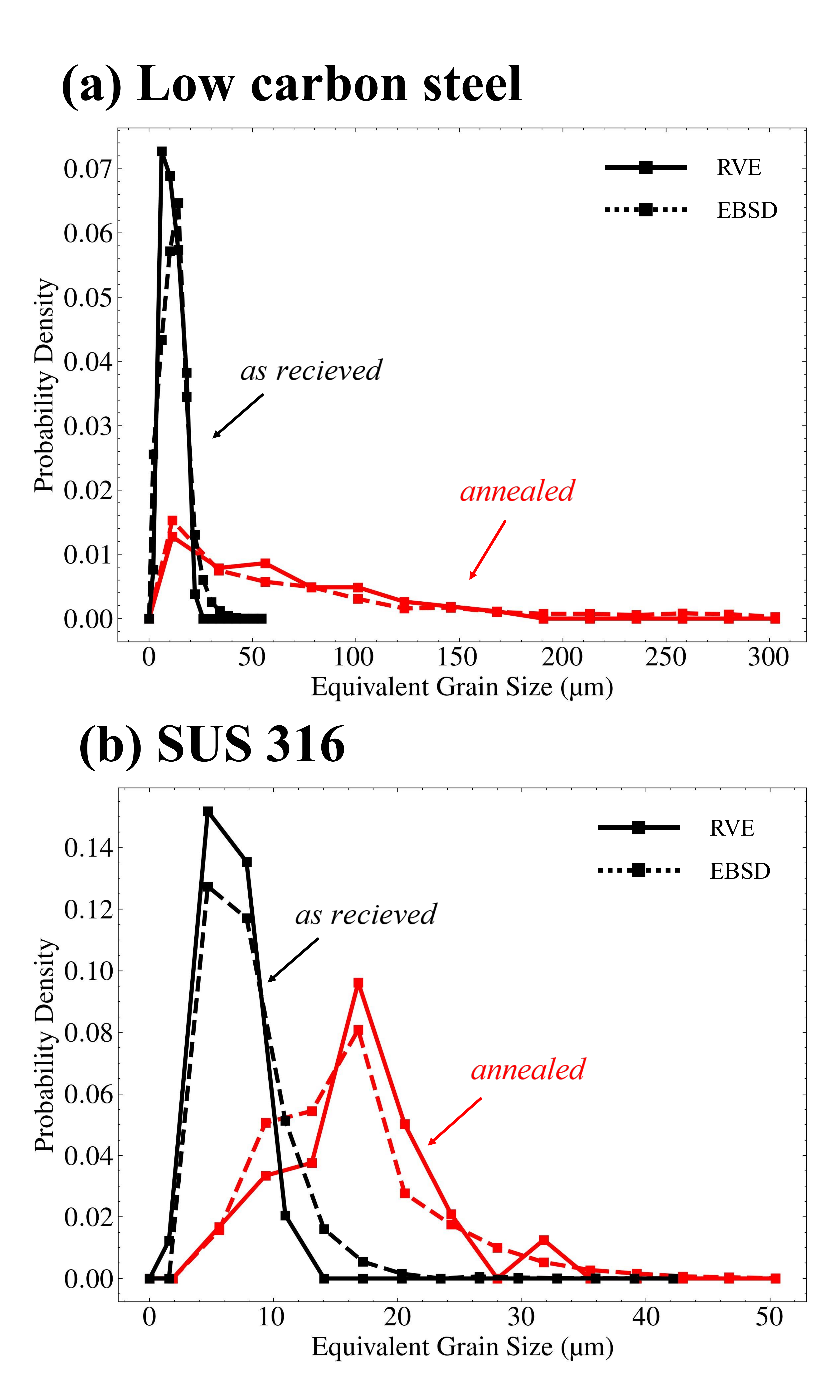}
\caption{Grain size distributions for (a) low-carbon steel and (b) SUS316, comparing experimental EBSD measurements (dotted lines) and the corresponding RVEs used in CPFEM simulations (solid lines). Both as-received (black) and annealed (red) conditions are shown. The RVE grain size distributions were generated to closely match the experimental EBSD data for each condition.}\label{fig:grain_size_distribution}
\end{figure}

\begin{figure}[!htbp]
\centering
\includegraphics[width=\textwidth]{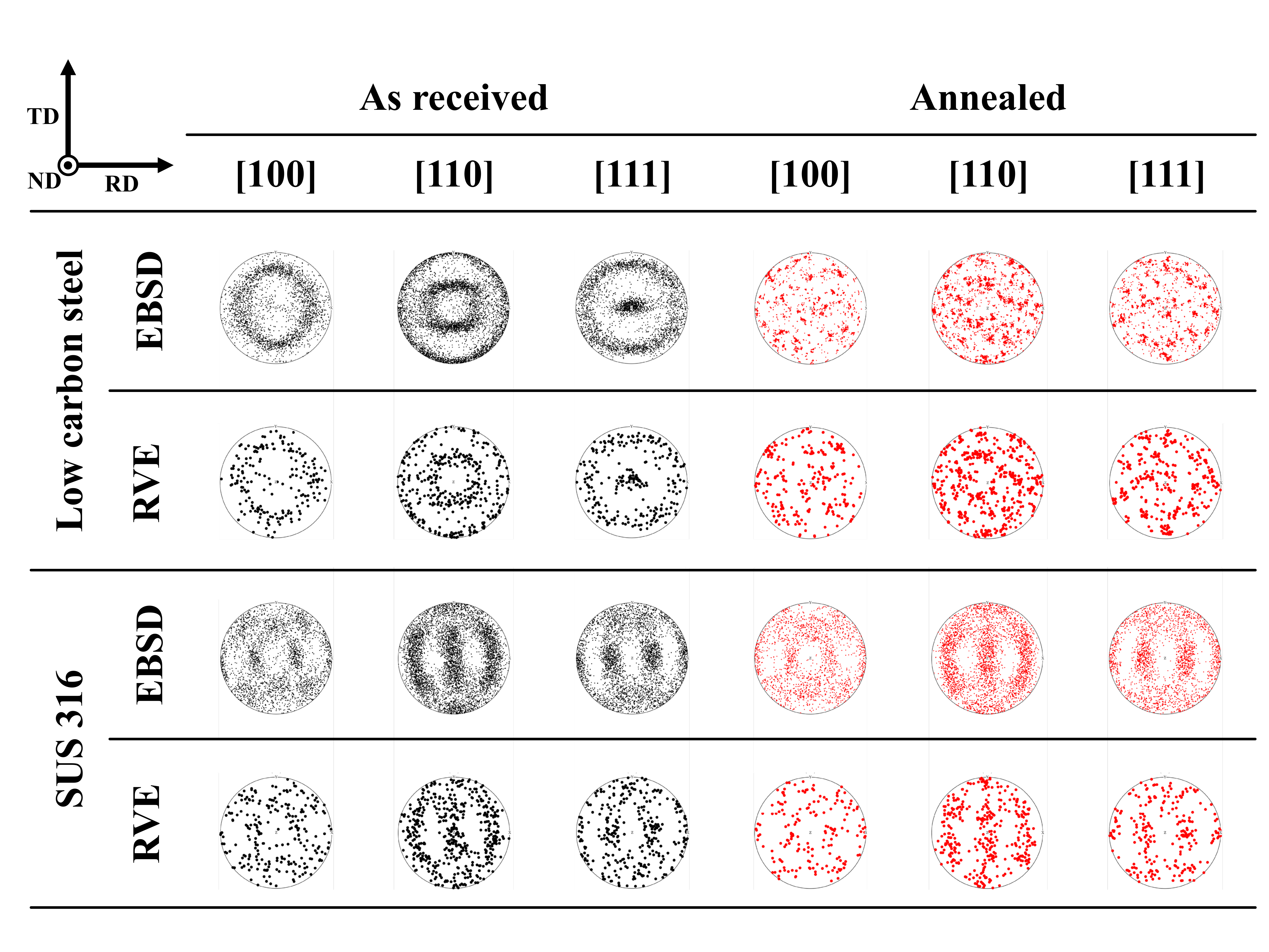}
\caption{Comparison of crystallographic texture between experiment and simulation. ODFs are shown for (top two rows) low-carbon steel and (bottom two rows) SUS316, in both as-received and annealed conditions. For each material and condition, experimental EBSD data (first and third rows) and corresponding RVEs used in CPFEM simulations (second and fourth rows) are presented. Each column represents a different sample direction: [100], [110], and [111], with the Rolling Direction (RD), Transverse Direction (TD), and Normal Direction (ND) indicated. The close agreement between EBSD and RVE ODFs demonstrates that the simulated microstructures accurately reproduce the experimentally measured textures.}\label{fig:ODFs}
\end{figure}

To represent these microstructures in CPFEM simulations, RVEs were constructed based on the EBSD data. The RVEs were generated to closely match the experimentally measured grain size distributions and crystallographic textures, as shown in \cref{fig:grain_size_distribution} and \cref{fig:ODFs}. The RVEs were meshed with $30^3$ elements, ensuring that the element size was one-tenth of the average grain size to avoid mesh-induced artifacts, as discussed in \cref{RVE and PBC}.

\begin{figure}[!htbp]
\centering
\includegraphics[width=\textwidth]{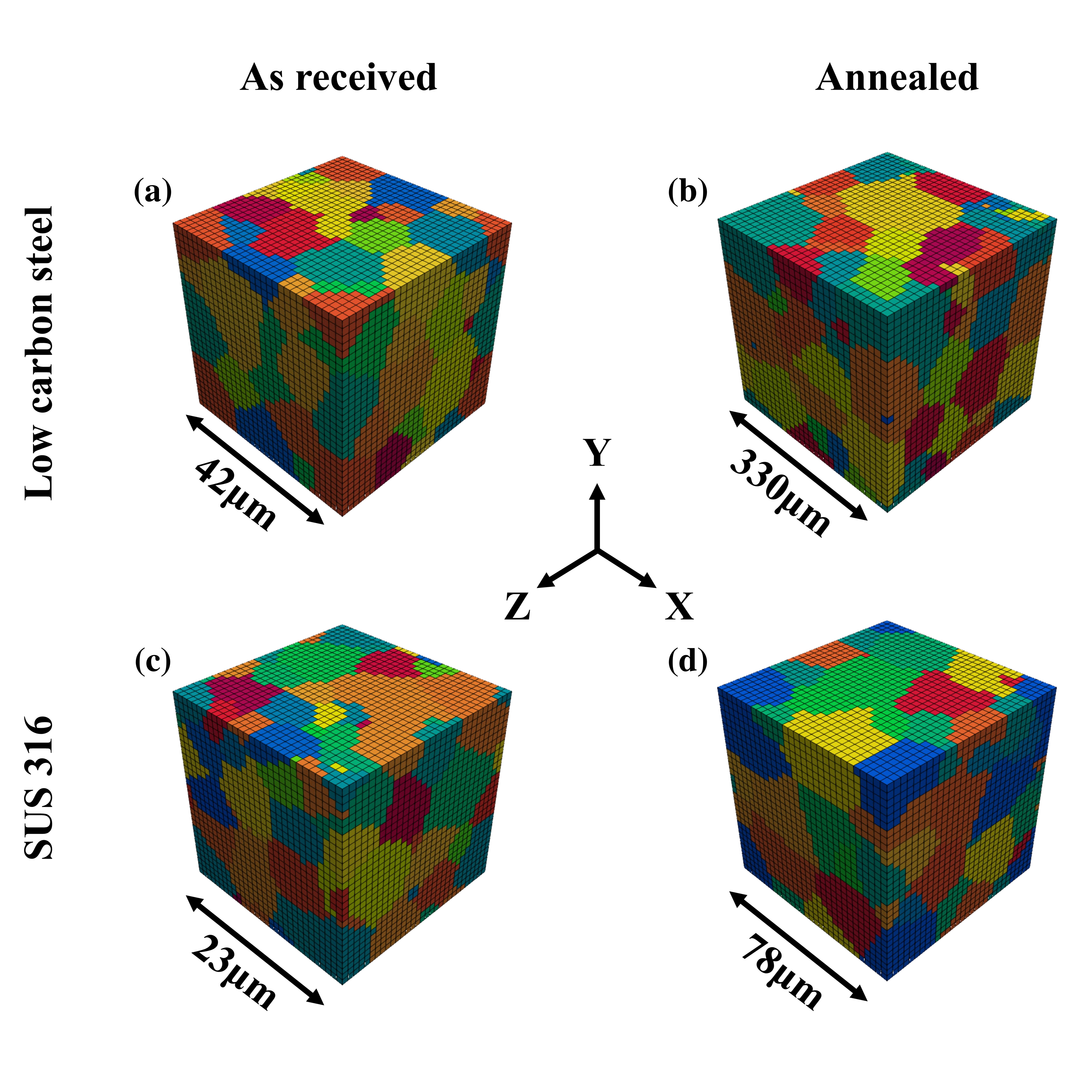}
\caption{RVEs constructed for CPFEM simulations, showing grain structures for (a) as-received low-carbon steel ($42~\mu$m), (b) annealed low-carbon steel ($330~\mu$m), (c) as-received SUS316 ($23~\mu$m), and (d) annealed SUS316 ($78~\mu$m). Each RVE size is set to ten times the average grain size to avoid mesh-induced artifacts, and all RVEs are meshed with $30^3$ elements.}\label{fig:RVEs}
\end{figure}

\subsection{Parameter calibration from tensile tests with different grain sizes}
\label{calibration}

In the proposed crystal-plasticity framework, the following 13 material parameters need to be identified: 
\[
\{\dot{\gamma}_0,\;m,\;\alpha,\;C_{11},\;C_{12},\;C_{44},\;k_1,\;k_2,\;(\rho_{\rm SSD}^a)_0,\;(\rho_{\rm GND}^a)_0,\;\gamma^{\rm sat},\;\zeta\}
\]
Here, $\dot{\gamma}_0$ is a reference shear rate scale fixed at $10^{-3}\,\mathrm{s}^{-1}$, and the strain rate sensitivity exponent $m$ is set to 0.05, following a previous study~\cite{Kalidindi1992}. The single-crystal elastic constants $C_{11}$, $C_{12}$, and $C_{44}$ for ferritic low-carbon steel and SUS316 are taken from Teodosiu's work~\cite{Teodosiu1982}. For simplicity, the initial GND density $(\rho_{\rm GND}^a)_0$ is set to zero (or can be set to a much smaller value than the commonly used SSD). The saturation shear increment $\gamma^{\rm sat}$, which has negligible influence under purely tensile loading, is fixed at $0.04$. Although $\alpha$ has a physical meaning rooted in dislocation interactions, it practically serves as a scaling factor for $\rho_{\rm SSD}^a$ in the stress-strain response. This interpretation is further discussed in Section~\ref{minor_parameters}. The remaining hardening parameters are $\{k_1,\;k_2,\;(\rho_{\rm SSD}^a)_0,\;\zeta\}$.

Because the initial statistically stored dislocation density \((\rho_{\rm SSD}^a)_0\) differs between annealed and as-received specimens, we first adjust these two initial SSD values manually to match the yield stress via \cref{eq:g}, which scales linearly with \((\rho_{\rm SSD}^a)_0\). The three parameters \(\{k_1,k_2,\zeta\}\) are then optimized simultaneously using the Nelder-Mead simplex algorithm~\cite{NelderMead1965} implemented in SciPy~\cite{Virtanen2020}. We collect the optimizable parameters into the vector $ \boldsymbol{\theta}=\{k_1,\;k_2,\;\zeta\}$ and minimize the average mean-square error between simulation and experiment over both specimen conditions:
\begin{equation}
\mathcal{L}(\boldsymbol{\theta})
= \frac{1}{2}\Bigl[\mathcal{L}_{\mathrm{ann}}(\boldsymbol{\theta})
               + \mathcal{L}_{\mathrm{ar}}(\boldsymbol{\theta})\Bigr],
\end{equation}
with
\begin{equation}
\mathcal{L}_{\mathrm{ann}}
= \frac{1}{N_{\mathrm{ann}}}
  \sum_{i=1}^{N_{\mathrm{ann}}}
  \bigl(\sigma_{\mathrm{sim}}^{\mathrm{ann}}(\varepsilon_i;\boldsymbol{\theta})
       - \sigma_{\mathrm{exp}}^{\mathrm{ann}}(\varepsilon_i)\bigr)^2,
\end{equation}
\begin{equation}
\mathcal{L}_{\mathrm{ar}}
= \frac{1}{N_{\mathrm{ar}}}
  \sum_{i=1}^{N_{\mathrm{ar}}}
  \bigl(\sigma_{\mathrm{sim}}^{\mathrm{ar}}(\varepsilon_i;\boldsymbol{\theta})
       - \sigma_{\mathrm{exp}}^{\mathrm{ar}}(\varepsilon_i)\bigr)^2.
\end{equation}

Here, \(\sigma_{\mathrm{sim}}^{\mathrm{ann}}\) and \(\sigma_{\mathrm{sim}}^{\mathrm{ar}}\) are the simulated true stresses at strain \(\varepsilon_i\) for the annealed and as-received conditions, respectively, while \(\sigma_{\mathrm{exp}}^{\mathrm{ann}}\) and \(\sigma_{\mathrm{exp}}^{\mathrm{ar}}\) are the corresponding experimental values. The number of data points in each condition is denoted by \(N_{\mathrm{ann}}\) and \(N_{\mathrm{ar}}\). 
The resulting parameters are summarized in \cref{tab:cpfem_all}. Note that \(k_1\), \(k_2\), and \(\zeta\) are shared between as-received and annealed conditions for each material, while the initial SSD density \((\rho_{\rm SSD}^a)_0\) is fitted separately for each condition.

\begin{table}[htbp]
\centering
\caption{Optimized CPFEM parameters for low-carbon steel and SUS316.}
\label{tab:cpfem_all}
\begin{tabular}{llcc}
\hline
\textbf{Category} & \textbf{Parameter} & \textbf{Low-carbon steel} & \textbf{SUS316} \\
\hline
\textit{Hardening} 
     & $k_1~[\mu\mathrm{m}]$      & 0.099   & 0.028 \\
     & $k_2~[-]$                  & 110     & 6.6   \\
     & $\zeta~[-]$                & 0.0098  & 0.033 \\
\midrule
\textit{Initial SSD density} 
     & $(\rho_{\rm SSD}^a)_0^{\rm ar}~[\mu\mathrm{m}^{-2}]$ & 1.6   & 7.1  \\
     & $(\rho_{\rm SSD}^a)_0^{\rm ann}~[\mu\mathrm{m}^{-2}]$ & 0.48  & 2.3  \\
\midrule
\textit{Elastic constants} 
     & $C_{11}~[\mathrm{GPa}]$    & 230.1   & 268.5 \\
     & $C_{12}~[\mathrm{GPa}]$    & 134.6   & 156.0 \\
     & $C_{44}~[\mathrm{GPa}]$    & 116.6   & 136.0 \\
     & $b~[\mathrm{nm}]$          & 0.248   & 0.252 \\
\midrule
\textit{Numerical}
     & $\alpha~[-]$               & 0.4     & 0.4   \\
     & $\Delta\gamma^{\rm sat}~[-]$ & 0.04    & 0.04  \\
     & $m~[-]$                    & 0.05    & 0.05  \\
     & $\dot{\gamma}_0~[\mathrm{s}^{-1}]$ & $1\times10^{-3}$ & $1\times10^{-3}$ \\
\hline
\end{tabular}
\end{table}

\begin{figure}[!htbp]
\centering
\includegraphics[width=0.6\textwidth]{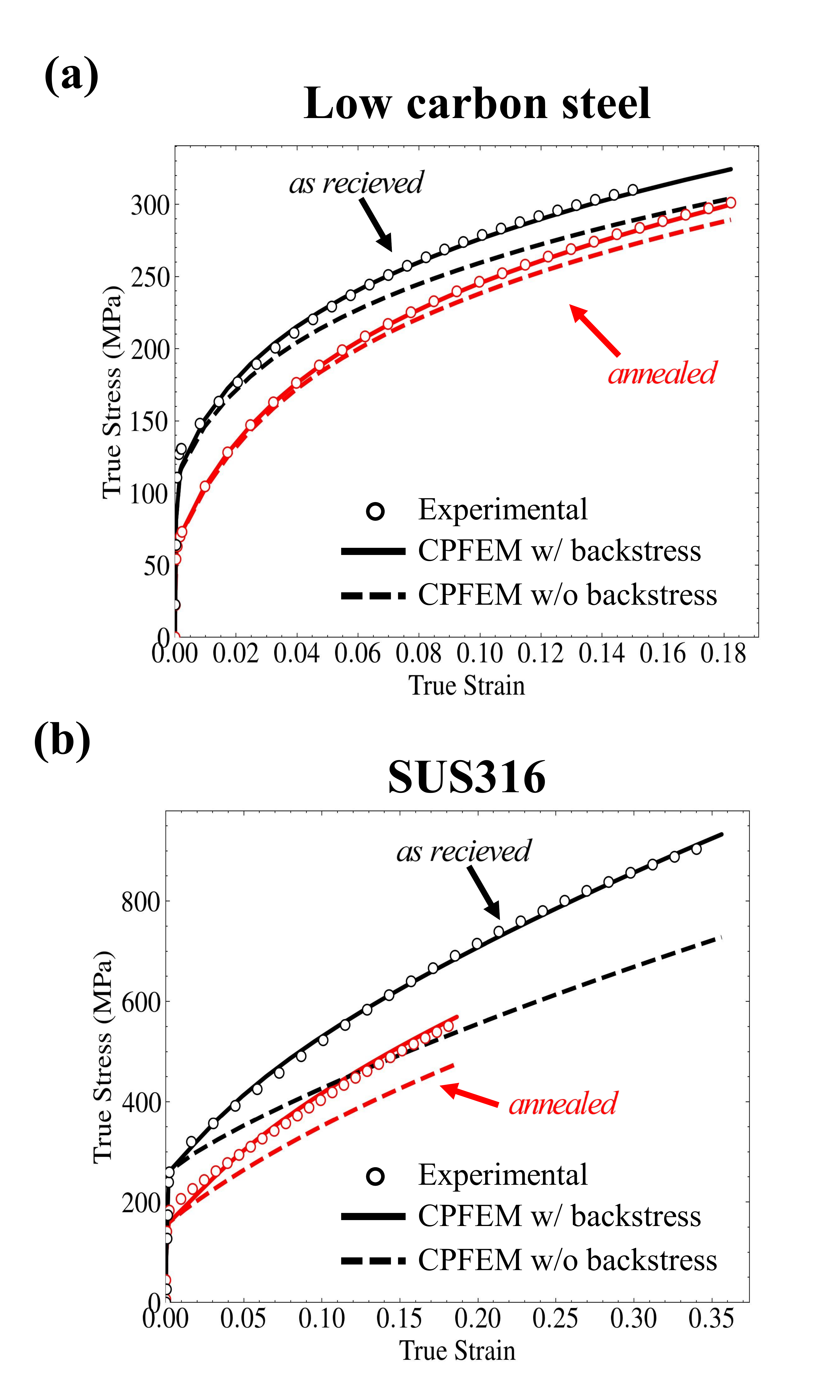}
\caption{Comparison of experimental and simulated true stress-strain curves for (a) low-carbon steel and (b) SUS316 under uniaxial tension, for both as-received (black) and annealed (red) conditions. Circles denote experimental data, solid lines show CPFEM simulations with the proposed GND-based back stress model, and dashed lines show CPFEM results without back stress.}\label{fig:optimized_sscurves}
\end{figure}

The resulting true stress-strain curves for both low-carbon steel and SUS316, comparing experimental data with CPFEM simulations, are shown in \cref{fig:optimized_sscurves}. The solid lines represent the CPFEM predictions using the calibrated parameters, while the dashed lines show the results from CPFEM simulations without the GND-based back stress for comparison (i.e., $\zeta=0$). The difference between the solid and dashed lines is more significant for the as-received conditions, indicating that the back stress is larger for smaller grains, as they possess a higher GND density. This makes the parameter $\zeta$ uniquely identifiable from the difference in stress-strain responses between the two microstructures.

\subsection{Validation: Tension-Compression (T-C) test}
\label{validation1}
The CPFEM model was validated by comparing the simulated tension-compression response of as-received low-carbon steel with experimental data from a tension-compression test. The tension-compression test was performed using the apparatus developed by Boger et al.~\cite{BOGER20052319}, as described in \cref{T-C test}.

\begin{figure}[!htbp]
\centering
\includegraphics[width=\textwidth]{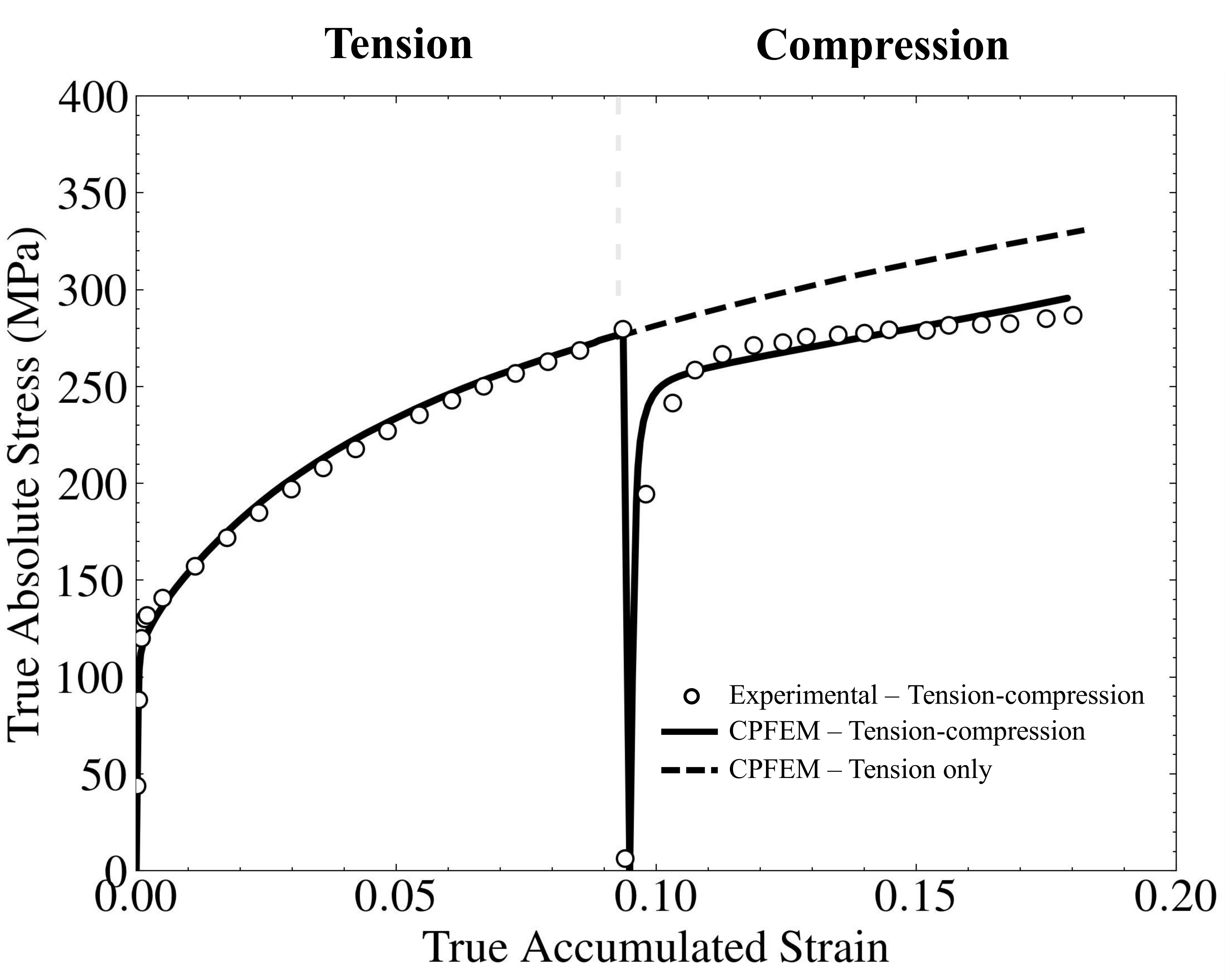}
\caption{Comparison of experimental and simulated tension-compression responses for as-received low carbon steel. Circles represent experimental data from the tension-compression test. The solid line shows the CPFEM prediction using parameters calibrated solely from uniaxial tensile tests on as-received and annealed specimens (i.e., without any tension-compression data). The dashed line is tensile result same as as-received low carbon steel RVE in \cref{fig:optimized_sscurves}.}\label{fig:TC_prediction}
\end{figure}

The tension-compression test results are shown in \cref{fig:TC_prediction}. The circles represent the experimental data from the tension-compression test, while the solid line shows the CPFEM prediction using parameters calibrated solely from uniaxial tensile tests on as-received and annealed specimens (i.e., without any tension-compression data). The dashed line represents the tensile result from the same as-received low-carbon steel RVE in \cref{fig:optimized_sscurves}.

The CPFEM prediction captures the tension-compression response well, demonstrating that the model can accurately predict the material behavior under cyclic loading conditions, even when calibrated only from uniaxial tensile tests.    

\subsection{Validation: Tension-Bending (T-B) test}
\label{validation2}
\begin{figure}[!htbp]
\centering
\includegraphics[width=\linewidth]{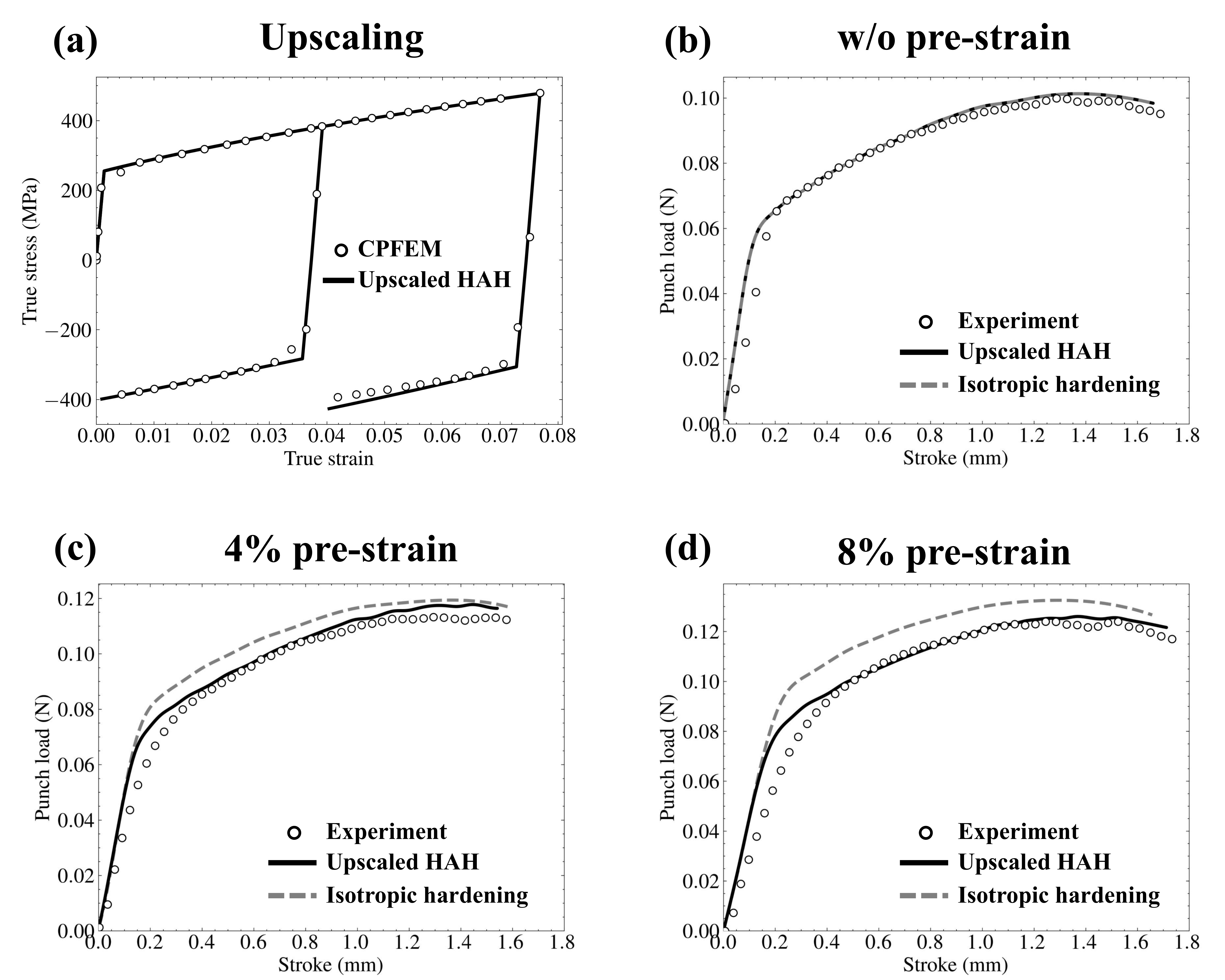}
\caption{Validation of the upscaled HAH model for tension-bending behavior. (a) Comparison of CPFEM and upscaled HAH model predictions for the tension-compression response of as-received SUS316. (b-d) Tension-bending test results for as-received SUS316: (b) without pre-strain, (c) with 4\% pre-strain, and (d) with 8\% pre-strain. Circles denote experimental punch load-stroke data, solid lines show predictions from the upscaled HAH model, and dashed lines indicate isotropic hardening predictions. The upscaled HAH model, calibrated only from grain size-dependent tensile data, outperforms isotropic hardening.}\label{fig:TB_prediction}
\end{figure}

The upscaled HAH model parameters were optimized to match CPFEM tension-compression responses for SUS316 at two specific pre-strain levels: 4\% and 8\%, as shown in \cref{fig:TB_prediction}(a). 
The optimization was performed using the Nelder-Mead simplex algorithm, minimizing the mean squared error between the HAH model and CPFEM stress-strain curves for 4\% pre-strain followed by 4\% compression, and 8\% pre-strain followed by 4\% compression.
These pre-strain levels correspond directly to the experimental tension-bending test conditions for SUS316, ensuring that the HAH model calibration is consistent with the subsequent experimental validation.
The optimized HAH parameters are summarized in \cref{tab:HAH_params}.

\begin{table}[htbp]
\centering
\caption{Optimized parameters for the upscaled HAH model for SUS316, obtained by fitting to CPFEM tension-compression curves at different pre-strain levels.}

\label{tab:HAH_params}
\begin{tabular}{cccc}
\hline
\textbf{Parameter} & \textbf{Description} & \textbf{Unit} & \textbf{Value} \\
$q$         & Yield surface distortion exponent     & [-] & 1.8  \\
$k$         & Microstructure deviator evolution rate & [-] & 0.12 \\
$k_g$       & Distortion evolution rate             & [-] & 0.09 \\
$\tau_g$    & Distortion saturation value           & [-] & 0.85 \\
$\gamma_g$  & Distortion saturation exponent        & [-] & 1.2  \\
$k_p$       & Permanent softening rate              & [-] & 0.04 \\
$\tau_p$    & Permanent softening saturation value  & [-] & 1.0  \\
\hline
\end{tabular}
\end{table}
The upscaled HAH model was then validated by comparing its predictions with experimental results from tension-bending tests on as-received SUS316. 
For force-displacement responses—\cref{fig:TB_prediction}(b-d)—the friction coefficient was first adjusted in \cref{fig:TB_prediction}(b) to match the experimental force-displacement curve, which was 0.15. The same coefficient was then used for \cref{fig:TB_prediction}(c) and \cref{fig:TB_prediction}(d). 

While the isotropic hardening model fails to capture the tension-bending behavior, the upscaled HAH model, calibrated only from grain size-dependent tensile data, outperforms isotropic hardening. This is because it accounts for the Bauschinger effect, which correctly estimates the inner side of the bending area, where the material is in compression after pre-straining in tension.

\section{Discussion}
\label{Discussion}

\subsection{Identifiability of the CPFEM model parameters}
\label{Identifiability of CPFEM model parameters}

Since the uniqueness of parameters from two annealed/as-received stress-strain curves is a key point of this research, we assess the identifiability of \(\{k_1,k_2,\zeta\}\) via a local sensitivity-based Fisher information analysis. Originally, identifiability was first used for parameter identification of viscoelastic models from nanoindentation tests by Richard et al.~\cite{RICHARD201341} and later expanded to CPFEM by Renner et al.~\cite{RENNER2020103916}. 

For each parameter \(\theta_j\in\{k_1,k_2,\zeta\}\) and each true-strain data point \(i\) (combining both annealed and as-received curves), the finite-difference, normalized sensitivity is calculated as:
\begin{equation}
S_{ij} =
\frac{\sigma_i(\theta_j + \xi\,\theta_j)
     - \sigma_i(\theta_j)}
     {\xi\,\theta_j}
\;\times\;
\frac{\theta_j}{\sigma_i(\theta_j)},
\end{equation}
where \(\xi=10^{-3}\) and \(\sigma_i\) is the simulation stress at the \(i\)th true-strain. Stacking all \(N\) values (corresponding to \(i\)) in the rows and the three parameters (corresponding to \(j\)) in the columns yields \(S\in\mathbb{R}^{N\times3}\), and the approximate Fisher information matrix is \(F = S^\mathsf{T} S\). Let \(\{\lambda_1\le\lambda_2\le\lambda_3\}\) be the eigenvalues of \(F\). The condition number is computed as \(\kappa = \lambda_{\max}/\lambda_{\min}\), and the average absolute sensitivity for each parameter is:

\begin{equation}
\delta_j = \frac{1}{N}\sum_{i=1}^N \bigl\lvert S_{ij}\bigr\rvert.
\end{equation}

For the combined (annealed and as-received) dataset, we obtain the following values:
The parameters $\lambda_1$, $\lambda_2$, and $\lambda_3$ are 0.4339, 8.7904, and 47.3934, respectively. The value of $\kappa$ is about 109.2. 
% Additionally, the parameters $\delta_{k_1}$, $\delta_{k_2}$, and $\delta_{\zeta}$ are 0.3625, 0.3107, and 0.0369, respectively. 
This result demonstrates that the parameters are uniquely identifiable (\(\kappa \approx 109.2\)), referring to Renner et al.~\cite{RENNER2020103916}'s criteria, where $log_{10}(\kappa)$ near or less than 2 indicates identifiability.

\subsection{Effect of $\Delta \gamma ^ {sat}$ and $\alpha$ on stress-strain curve}
\label{minor_parameters}

\begin{figure}[!htbp]
\centering
\includegraphics[width=0.8\textwidth]{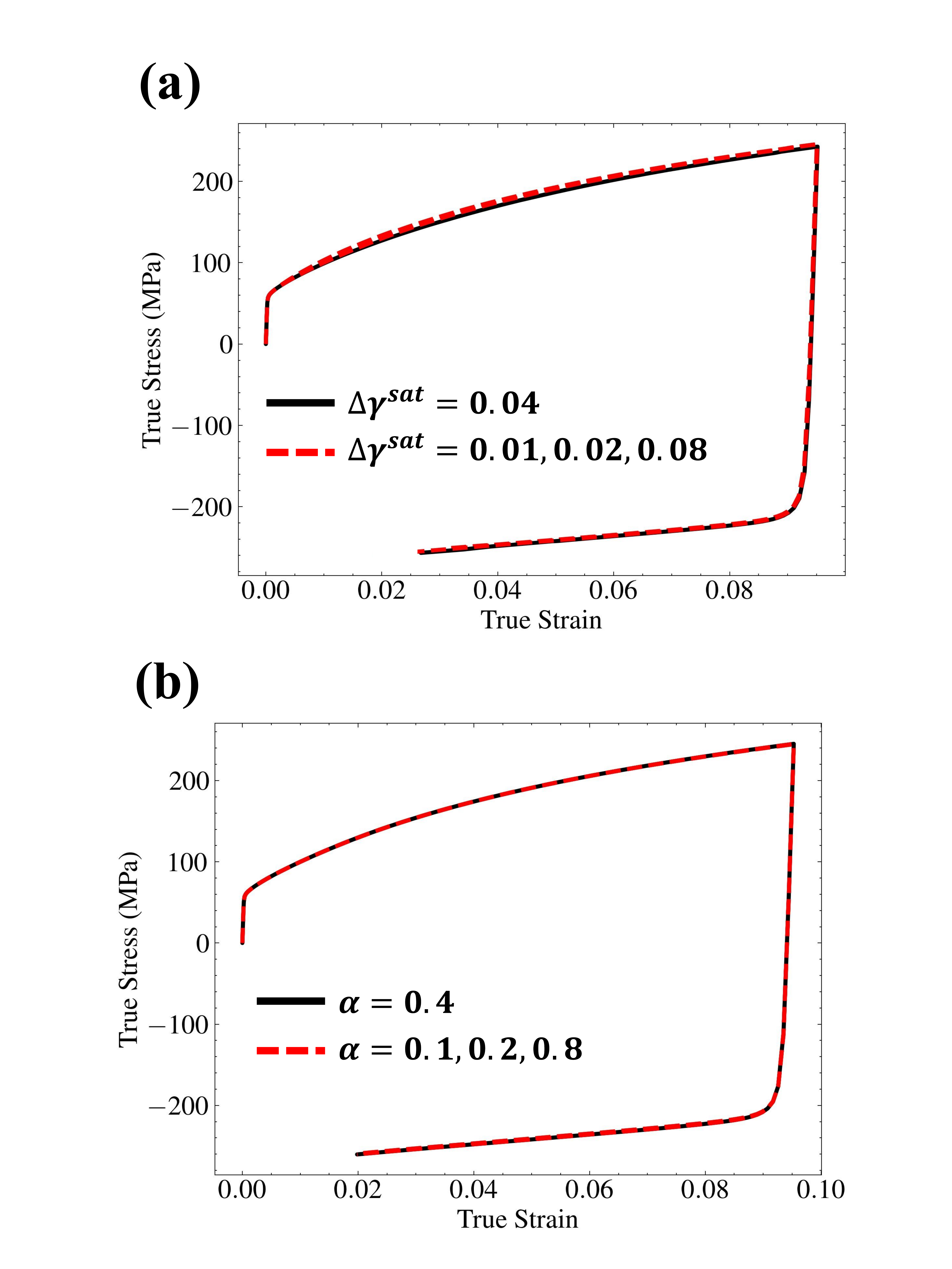}
\caption{Sensitivity analysis of minor model parameters on the simulated tension-compression stress-strain response for low-carbon steel. (a) Effect of the slip saturation threshold $\Delta\gamma^{\mathrm{sat}}$ on the stress-strain curve. Varying $\Delta\gamma^{\mathrm{sat}}$ over a wide range (0.01-0.08) has negligible influence on the response. (b) Effect of the geometric factor $\alpha$ in the Taylor hardening law. The stress-strain response is nearly insensitive to $\alpha$ within the tested range (0.1-0.8), as it primarily acts as a scaling factor for the initial SSD density.}\label{fig:minor_parameters}
\end{figure}

In the current CPFEM model, the slip saturation threshold \(\Delta\gamma^{\mathrm{sat}}\) and the geometric factor \(\alpha\) in the Taylor hardening law are considered to have a minor influence. For \(\Delta\gamma^{\mathrm{sat}}\) in \cref{eq:back stress}, it does not have a strong effect after slip saturation, which is typically reached at low strains. Therefore, we varied \(\Delta\gamma^{\mathrm{sat}}\) over a wide range (0.01-0.08) and found that it has negligible influence on the stress-strain response, as shown in \cref{fig:minor_parameters}(a).

Similarly, the geometric factor \(\alpha\) in \cref{eq:g}, which originally has a physical meaning rooted in dislocation interactions, acts primarily as a scaling factor for the initial SSD density \((\rho_{\rm SSD}^a)_0\). In fact, \cref{eq:g}, \cref{eq:rho_forest}, and \cref{eq:KM1} become equivalent when \(\alpha\) is enlarged by a factor of \(c\), \((\rho_{\rm SSD}^a)_0\) is reduced by the same factor \(1/c^2\), and \(k_1\) is scaled by \(1/c\). Therefore, \(\alpha\) does not affect the stress-strain response, as it only scales the initial SSD density and the hardening parameters.

From the above analysis, 
we varied \(\alpha\) over a range of 0.1-0.8, scaling \((\rho_{\rm SSD}^a)_0\) and \(k_1\) accordingly, and found that the stress-strain responses are virtually identical, as shown in \cref{fig:minor_parameters}(b). This confirms that \(\alpha\) primarily serves as a scaling factor for the initial SSD density and does not significantly influence the stress-strain response.

\subsection{Work hardening stagnation and permanent softening}
\label{shear_reverse_shear}

Shear-reverse shear tests are widely used to investigate mechanical response under load reversal, as they are experimentally more accessible than tension-compression tests~\cite{YOSHIDA2002661,BARLAT20111309,BARLAT2013130,RAUCH20072939}. To compare the proposed model with previous studies, we performed shear-reverse shear simulations with PBC on the as-received low-carbon steel RVE, following~\cite{BARLAT20111309,RAUCH20072939}.

A standard practice in analyzing shear-reverse shear results is to horizontally shift the shear-reverse shear stress-strain curve to the left so it overlaps with the monotonic shear curve. The amount of shift required for the curves to overlap is called Bauschinger strain, which Rauch et al.~\cite{RAUCH20072939} showed matches the amount of prestrain. They further explained that work-hardening stagnation and permanent softening after reversal are caused by the dissolution of cell structures formed during prestrain. To model these phenomena, previous studies introduced phenomenological variables such as forward/reverse dislocation density~\cite{RAUCH20072939}.

\begin{figure}[!htbp]
\centering
\includegraphics[width=\textwidth]{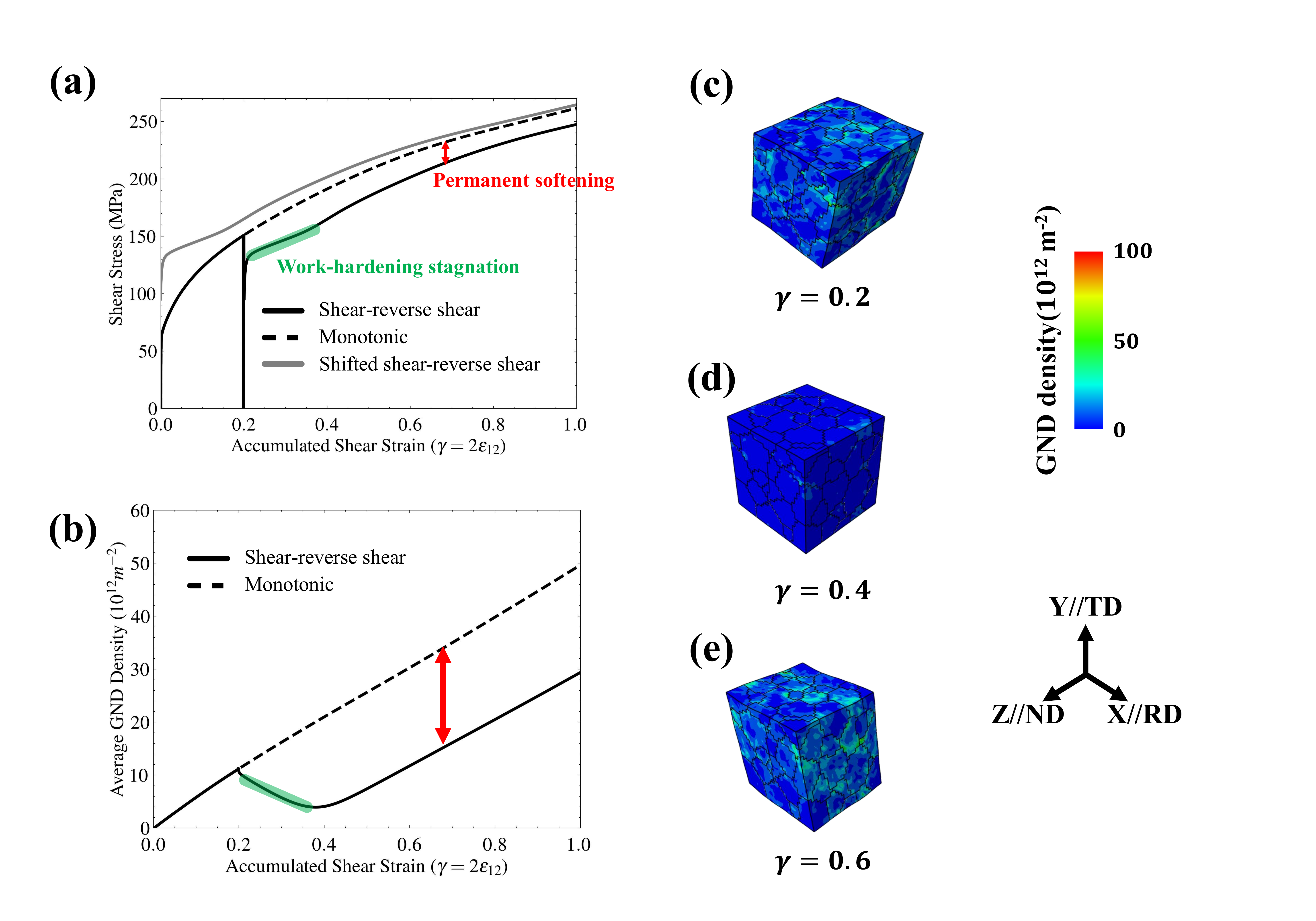}
\caption{
Shear-reverse shear simulation results for the as-received low-carbon steel RVE ((c) in \cref{fig:RVEs}).
(a) Shear stress-shear strain curves for monotonic shear loading and shear-reverse shear loading.
(b) Evolution of GND density during monotonic shear and shear-reverse shear loading.
(c-e) GND density distributions at key stages:
(c) just before reversal (accumulated shear strain 0.2),
(d) after reversal (accumulated shear strain 0.4, shear strain 0.0),
(e) after permanent softening (accumulated shear strain 0.6, shear strain -0.2).
The color scale indicates GND density.
Coordinate axes: Rolling Direction (RD), Transverse Direction (TD), and Normal Direction (ND).
}\label{fig:shear_reverse_shear}
\end{figure}

Our simulation results are shown in \cref{fig:shear_reverse_shear}. In (a), the dashed line represents the reverse shear segment shifted left by 0.2. The overlap between the shifted reverse shear and monotonic shear curves confirms that the Bauschinger strain equals the prestrain, consistent with Rauch et al.~\cite{RAUCH20072939}. The region with reduced slope during reversal indicates work-hardening stagnation, and the lower stress in the reverse direction compared to monotonic loading demonstrates permanent softening.

The underlying mechanism in our model is explained by the evolution of GND density, shown in panel (b). Unlike previous phenomenological approaches, our model computes GND density directly from the geometric incompatibility in the finite element mesh. As shown in panels (c)-(e), GND density increases during forward shear (c), decreases during reversal as the lattice curvature relaxes (d), and increases again during continued reverse shear (e). The reduction in GND density during reversal corresponds to the dissolution of cell structures, which naturally produces work-hardening stagnation and permanent softening without additional empirical terms. Even as GND density increases again in the reverse direction, it remains lower than in monotonic loading, explaining the persistent softening in (a).

Thus, the proposed physically motivated GND-based back stress model reproduces key features of load reversal behavior—Bauschinger strain, work-hardening stagnation, and permanent softening—through explicit microstructural evolution, rather than phenomenological fitting.

% \subsection{Influence of inclusions}
% \label{Influence of inclusions}

% Subsection text

\section{Conclusions}
\label{Conclusions}

In this study, we developed a grain size-informed hardening inversion methodology for ultra-thin metal sheets, enabling the prediction of tension-compression and strain-path change behavior using only uniaxial tensile data from specimens with different grain sizes. The key findings are as follows:

\begin{itemize}
     \item We demonstrated that the explicit superposition of PK forces between integration points in CPFEM leads to conceptual inconsistencies, computational inefficiency, and redundancy, as CPFEM already implicitly accounts for elastic dislocation interactions.
     \item A GND-based back stress model was proposed, physically motivated by the formation of GNBs and incorporating crystallographic alignment between slip systems. This model introduces a single parameter, $\zeta$, which can be uniquely identified from tensile tests on specimens with different grain sizes.
     \item The CPFEM model, calibrated only from tensile data, accurately predicted tension-compression responses and captured the Bauschinger effect in low-carbon steel, as validated by direct experiments.
     \item For ultra-thin SUS316 sheets where direct compression testing is infeasible, the upscaled HAH model—calibrated from virtual CPFEM tension-compression simulations—successfully validated tension-bending behavior.
     \item Sensitivity and identifiability analyses confirmed that the key model parameters are uniquely determined by the available experimental data, and the model robustly captures permanent softening and work-hardening stagnation under strain path changes.
\end{itemize}

Overall, the proposed methodology provides a practical and physically motivated framework for predicting complex cyclic and reverse loading behavior in thin metal sheets, using only accessible tensile test data and microstructural information.

In this era of machine learning and data-driven approaches, our work emphasizes the importance of physical understanding and model interpretability in materials modeling. Machine learning methods can complement and enhance traditional modeling approaches~\cite{HU2024113121,SIM2025116407,WHITMAN2025121217}, but it is difficult to predict beyond the data without a solid physical foundation. 
Our grain size-informed hardening inversion methodology provides a robust framework for predicting beyond the data for parameter identification, enabling the prediction of complex material behavior under tension-compression using only uniaxial tensile data from specimens with different grain sizes. This approach highlights the value of combining physical insights with computational methods to advance materials modeling and simulation.
 
% \appendix
% \section{Example Appendix Section}


% \section{Example Appendix Section}
% \label{app2}


\bibliography{cas-refs.bib}

\section{Acknowledgements}

\end{document}